\documentclass[a4paper,12pt]{article}
\usepackage{ascmac}
\usepackage{color}
\newcommand{\BQED}{\hfill \hbox{\rule{8pt}{8pt}}}

\newenvironment{namelist}[1]{%
\begin{list}{}
  { 
	\settowidth{\labelwidth}{#1}
	\setlength{\leftmargin}{1.1\labelwidth}}
  \setlength{\itemsep}{0cm}
}{% 
\end{list}}
\catcode`\@=11
\def\newexample#1{\@ifnextchar[{\@oexm{#1}}{\@nexm{#1}}}

\def\@nexm#1#2{%
\@ifnextchar[{\@xnexm{#1}{#2}}{\@ynexm{#1}{#2}}}

\def\@xnexm#1#2[#3]{\expandafter\@ifdefinable\csname #1\endcsname
{\@definecounter{#1}\@addtoreset{#1}{#3}%
\expandafter\xdef\csname the#1\endcsname{\expandafter\noexpand
  \csname the#3\endcsname \@exmcountersep \@exmcounter{#1}}%
\global\@namedef{#1}{\@exm{#1}{#2}}\global\@namedef{end#1}{\@endexample}}}

\def\@ynexm#1#2{\expandafter\@ifdefinable\csname #1\endcsname
{\@definecounter{#1}%
\expandafter\xdef\csname the#1\endcsname{\@exmcounter{#1}}%
\global\@namedef{#1}{\@exm{#1}{#2}}\global\@namedef{end#1}{\@endexample}}}

\def\@oexm#1[#2]#3{\expandafter\@ifdefinable\csname #1\endcsname
  {\global\@namedef{the#1}{\@nameuse{the#2}}%
\global\@namedef{#1}{\@exm{#2}{#3}}%
\global\@namedef{end#1}{\@endexample}}}

\def\@exm#1#2{\refstepcounter
    {#1}\@ifnextchar[{\@yexm{#1}{#2}}{\@xexm{#1}{#2}}}

\def\@xexm#1#2{\@beginexample{#2}{\csname the#1\endcsname}\ignorespaces}
\def\@yexm#1#2[#3]{\@opargbeginexample{#2}{\csname
       the#1\endcsname}{#3}\ignorespaces}

\def\@exmcounter#1{\noexpand\arabic{#1}}
\def\@exmcountersep{.}
\def\@beginexample#1#2{\trivlist \item[\hskip 
\labelsep{\bf #1\ #2:}]}
\def\@opargbeginexample#1#2#3{\trivlist
      \item[\hskip \labelsep{\bf #1\ #2\ }#3{\bf :}]}
\def\@endexample{\endtrivlist}

\catcode`\@=12

\newtheorem{theorem}{{\bf Theorem}}[section]
\newtheorem{definition}{{\bf Definition}}[section]
\newtheorem{corollary}{{\bf Corollary}}[section]

\newtheorem{example}{{\bf Example}}[section]

\newcommand{\msc}[1]{\mbox{{\sc #1}}}
\catcode`\@=11

\def\@xnthm#1#2[#3]{\expandafter\@ifdefinable\csname #1\endcsname
{\@definecounter{#1}\@addtoreset{#1}{#3}%
\expandafter\xdef\csname the#1\endcsname{\expandafter\noexpand
\bf \csname the#3\endcsname \@thmcountersep \@thmcounter{#1}}%
\global\@namedef{#1}{\@thm{#1}{#2}}\global\@namedef{end#1}{\@endtheorem}}}

\def\@ynthm#1#2{\expandafter\@ifdefinable\csname #1\endcsname
{\@definecounter{#1}%
\expandafter\xdef\csname the#1\endcsname{\bf \@thmcounter{#1}}%
\global\@namedef{#1}{\@thm{#1}{#2}}\global\@namedef{end#1}{\@endtheorem}}}

\def\@begintheorem#1#2{\trivlist \item[\hskip 
\labelsep{\bf #1~#2:}]\sl}
\def\@opargbegintheorem#1#2#3{\trivlist
      \item[\hskip \labelsep{\bf #1~#2}~#3{\bf :}]\sl}

\catcode`\@=12

\newbox\rubisita
\newbox\rubiue
\newdimen\rubiw
\def\rubi#1#2{{\setbox\rubisita=\hbox{#1}\setbox\rubiue=\hbox{\tiny #2}%
\ifdim \wd\rubisita>\wd\rubiue\rubiw=\wd\rubisita\else\rubiw=\wd\rubiue\fi%
\kanjiskip=0pt plus1fil%
\setbox\rubisita=\hbox to \rubiw{\hfil#1\hfil}%
\setbox\rubiue=\hbox to \rubiw{\tiny\hfil#2\hfil}%
\vbox{\offinterlineskip\box\rubiue\break\box\rubisita}}}
\begin{document}
\begin{center}
{\Large {\bf Optimal Online Algorithms\smallskip\\ 
for the Multi-Objective Time Series Search Problem}}\bigskip\\
\begin{tabular}{ccc}
{\sc Shun Hasegawa} & & {\sc Toshiya Itoh}\\
{\sf hasegawa.s.aj@m.titech.ac.jp} & & {\sf titoh@ip.titech.ac.jp}\\
Department of Computer Science & & Department of Information Processing\\
Tokyo Institute of Technology & & Tokyo Institute of Technology
\end{tabular}
\end{center}\medskip
{\sf Abstract:} Tiedemann, et al. 
[Proc. of WALCOM, LNCS 8973, 2015, pp.210-221] defined 
multi-objective online problems 
and the 
competitive analysis for multi-objective online problems,~and 
showed best possible online algorithms with respect to several 
measures of the competitive~analysis. 
In this paper, we first point out that the definitions and 
frameworks~of~the~competitive~a\-nalysis due to Tiedemann, et al. 
do not necessarily capture the efficiency of 
online algorithms for multi-objective 
online problems and provide modified definitions of 
the competitive analysis 
for multi-objective online problems. 
Under the modified framework, we present a simple online algorithm 
Balanced Price Policy ($\msc{bpp}_{k}$) for the multi-objective 
($k$-objective) time series search problem, 
and show that the 
algorithm $\msc{bpp}_{k}$ is 
{\it best possible\/} with respect to any measure of the competitive analysis 
(defined by a monotone function $f$). 
For the modified framework, we also derive 
best possible values of the competitive ratio for the multi-objective 
time series search problem with respect to several representative 
measures of the competitive analysis. \medskip

\section{Introduction} \label{sec-introduction}
%==============================================
%
Single-objective online optimization problems are fundamental  in 
computing, communicating, and other practical 
systems. To measure 
the efficiency of online algorithms for single-objective 
online optimization problems, 
a notion of competitive analysis was introduced by Sleator and Tarjan \cite{ST}, 
and since then extensive research has been made for diverse areas, e.g.,  
paging~and caching (see \cite{Y} for a survey), 
metric task systems (see \cite{K} for a survey), 
asset conversion problems (see \cite{MAS} for a survey), 
buffer management of network switches (see \cite{G} for a survey), etc. 
All of these are single-objective online problems. 
In practice, there are many online problems of 
multi-objective nature, but we have no general framework of 
competitive analysis and no definition of competitive ratio for 
multi-objective online problems. 
Tiedemann, et al. \cite{TIS} first introduced a framework of 
multi-objective online problems as the online version of multi-objective 
optimization problems \cite{E} and formulated 
a notion of the competitive ratio for multi-objective online problems 
by extending the competitive ratio for single-objective 
online~problems.~To~define the competitive ratio for multi-objective ($k$-objective) 
online problems, Tiedemann,~et~al. \cite{TIS} regarded multi-objective online 
problems as a family of (possibly dependent) 
single-objective online problems and applied a monotone 
function $f:{\bf R}^{k}\to {\bf R}$~to~the~family of the single-objective online problems. Given an algorithm 
{\sc alg} for a multi-objective ($k$-objective) online problem, 
we regard {\sc alg} as a family of algorithms $\msc{alg}_{i}$ 
for the $i$th objective of the input~sequence and let $c_{i}$ be 
the competitive ratio of the algorithm $\msc{alg}_{i}$. 
For the set $\{c_{1},\ldots.c_{k}\}$~of~$k$~competitive ratios, 
the algorithm {\sc alg} is $f(c_{1},\ldots,c_{k})$-competitive 
with respect to a monotone function 
$f: {\bf R}^{k} \to {\bf R}$. 
In fact, Tiedemann, et al. \cite{TIS} defined the worst component 
competitive ratio by a function $f_{1}(c_{1},\ldots,c_{k}) = 
\max(c_{1},\ldots,c_{k})$, the arithmetic mean component 
competitive ratio by a function 
$f_{2}(c_{1},\ldots,c_{k})= 
(c_{1}+\cdots+c_{k})/k$, and 
the geometric mean 
component competitive ratio 
by a function $f_{3}(c_{1},\ldots,c_{k}) = 
(c_{1}\times \cdots \times c_{k})^{1/k}$. 
Note that all of the functions~$f_{1}$, $f_{2}$, and $f_{3}$ are continuous on 
${\bf R}^{k}$ and monotone. 
%
%=====================================================
\subsection{Previous Work} \label{subsec-previous}
%=====================================================
%
El-Yaniv, et al. \cite{Eetal} initially investigated the single-objective time 
series search problem. For the single-objective time 
series search problem, prices are revealed time by time and
the goal of the algorithm is to select one of them as with high price 
as possible. Assume that 
$m>0$ and $M>m$ are the minimum and maximum values of possible prices, 
respectively, and let $\phi=M/m$ be the {\it fluctuation ratio\/}  
of possible prices. 
Under the assumption that $M>m>0$~are~known~to online algorithms, 
El-Yaniv, et al. \cite{Eetal} presented a deterministic 
algorithm {\it reservation price policy\/} {\sc rpp}, 
which is  shown to be $\sqrt{\phi}$-competitive and best possible, 
and a randomized algorithm {\it exponential threshold\/} 
{\sc expo}, which is shown to be $O(\log \phi)$-competitive. 

In a straightforward manner, 
Tiedemann, et al. \cite{TIS} generalized 
the single-objective~time~series search problem and 
defined the multi-objective time series search problem. 
For the multi-objective ($k$-objective) time series search problem, 
a vector $\vec{p}=(p_{1},\ldots,p_{k})$ of 
$k$~(possibly~dependent) prices are revealed time by time 
and the goal of the algorithm is to select one of the price vectors 
as with low competitive ratio as possible with respect to the monotone 
function $f: {\bf R}^{k} \to {\bf R}$. 
For each $1 \leq i \leq k$, assume that $m_{i}>0$ and $M_{i}>m_{i}$ are 
the minimum and maximum values of possible prices for the $i$th objective, 
respectively, and $m_{i},M_{i}$ are known to online algorithms. 
For each $i \in [1,k]$, we use $\msc{itv}_{i}=[m_{1},M_{i}]$ to denote 
an interval of the prices for the $i$th objective. 
For the case that all of $\msc{itv}_{1}=[m_{1},M_{1}],\ldots,
\msc{itv}_{k}=[m_{k},M_{k}]$~are~{\it real\/}~intervals, 
Tiedemann, et al. \cite{TIS} presented best possible 
online algorithms for the multi-objective~time~series search problem with 
respect to the monotone functions 
$f_{1}$, $f_{2}$, and $f_{3}$, i.e., a best possible online algorithm 
for the multi-objective ($k$-objective) 
time series search 
problem with respect to the monotone function $f_{1}$ 
\cite[Theorems 1 and 2]{TIS}, a best possible online algorithm 
for the bi-objective time series search 
problem with respect to the monotone function 
$f_{2}$ \cite[Theorems 3 and 4]{TIS} and a best possible online algorithm 
for the bi-objective time series search problem with respect to the 
monotone function $f_{3}$ \cite[\S3.2]{TIS}. 
Note that the proofs of these results are correct under 
the assumption that 
all of $\msc{itv}_{1}=[m_{1},M_{1}],\ldots,
\msc{itv}_{k}=[m_{k},M_{k}]$ are {\it real\/} intervals. 
%
%===========================================================
\subsection{Our Contribution} \label{subsec-contribution}
%===========================================================
%
We first observe that the definition and framework 
of competitive analysis given by Tiedemann, et al. \cite[Definitions 1, 2, and 3]{TIS} 
do not necessarily capture the efficiency of algorithms for 
multi-objective online problems. Then we 
introduce modified 
definition and framework of competitive analysis 
for multi-objective online problems. 

As mentioned in Subsection \ref{subsec-previous}, 
Tiedemann, et al. \cite{TIS} showed best possible 
online algorithms for the multi-objective time series search problem with 
respect to the monotone continuous functions 
$f_{1}$, $f_{2}$ and $f_{3}$ under the assumption that all of 
$\msc{itv}_{1}=[m_{1}, M_{1}],\ldots,\msc{itv}_{k}=[m_{k},M_{k}]$~are 
real intervals, 
however, the optimality for the algorithm with respect to each of 
the monotone continuous 
functions $f_{1}$, $f_{2}$ and $f_{3}$ is discussed separately and 
independently. 
In this paper,~we present a simple online 
algorithm Balanced Price Policy ($\msc{bpp}_{k}$)  
for the multi-objective time series search problem with respect 
to any monotone 
function $f:{\bf R}^{k} \to {\bf R}$ and then show 
that under the modified framework of competitive analysis, 
the algorithm $\msc{bpp}_{k}$ is 
{\it best possible\/} for any monotone (not necessarily continuous) 
function $f:{\bf R}^{k} \to {\bf R}$~even~if~all~of~$\msc{itv}_{1}=[m_{1},M_{1}],\ldots,\msc{itv}_{k}=[m_{k},M_{k}]$ 
are not necessarily real intervals 
(in Theorem \ref{thm-general}). 
In the case that 
all of $\msc{itv}_{1}=[m_{1},M_{1}],\ldots,\msc{itv}_{k}=[m_{k},M_{k}]$ 
are real intervals, we exactly formulate~the competitive ratio of 
the algorithm $\msc{bpp}_{k}$ for any monotone function 
$f: {\bf R}^{k}\to {\bf R}$~(in~Theorems \ref{thm-unified-upper} 
and \ref{thm-unified-lower}). 
With respect to the {\it existing\/} monotone continuous functions 
$f_{1}$,~$f_{2}$,~and~$f_{3}$,~we~derive the best possible values 
of the competitive ratio 
for the multi-objective time series search problem 
under the modified framework of competitive analysis 
in Theorems~\ref{thm-worst},~\ref{thm-arithmetic},~and 
\ref{thm-geometric}, respectively. 
With respect to a {\it new\/} monotone function 
$f_{4}(c_{1},\ldots,c_{k})=\min(c_{1},\ldots,c_{k})$, we also 
derive the best possible value of the competitive ratio 
for the multi-objective time series search problem 
under the modified framework of competitive analysis 
in Theorem \ref{thm-best}. 

From Theorems \ref{thm-unified-upper} and \ref{thm-unified-lower}, 
we note that (1) Theorem \ref{thm-worst} 
gives another proof for the result that the algorithm in \cite[Theorem 1]{TIS} 
is best possible 
for the multi-objective time series search problem with respect to $f_{1}$, 
(2) Theorem \ref{thm-arithmetic} disproves the result that 
the algorithm in \cite[Theorem 3]{TIS} is best possible 
for the bi-objective time series search problem with 
respect to $f_{2}$, and (3) Theorem \ref{thm-geometric} gives a best possible 
online algorithm for the multi-objective time series search problem with respect 
to $f_{3}$, which is an extension of 
the result that the algorithm in 
\cite[Theorem 3]{TIS} is best possible for the bi-objective 
time series search problem with respect to $f_{3}$. 
%
%==============================================
\section{Preliminaries} \label{sec-preliminary}
%==============================================
%
For the subsequent discussions, we present some notations and terminologies. 
For any  pair~of~integers $a\leq b$, we use $[a,b]$ to denote 
a set $\{a,\ldots,b\}$ and for any pair of vectors 
$\vec{x} = (x_{1},\ldots,x_{k}) \in {\bf R}^{k}$ and $\vec{y}=
(y_{1},\ldots,y_{k}) \in {\bf R}^{k}$, we use 
$\vec{x}\preceq \vec{y}$ to 
denote a componentwise order, i.e., $x_{i}\leq y_{i}$~for each $i \in [1,k]$. 
It is immediate that $\preceq$ is a partial order on ${\bf R}^{k}$. 
A function $f: {\bf R}^{k}\to {\bf R}$ is said to be 
{\it monotone\/} if $f(\vec{x}) \leq f(\vec{y})$ for any pair of vectors 
$\vec{x} \in {\bf R}^{k}$ and $\vec{y} \in {\bf R}^{k}$ such that 
$\vec{x} \preceq \vec{y}$. 
%
%==============================================
\subsection{Multi-Objective Online Problems}
\label{subsec-multi-objective}
%==============================================
%
Tiedemann, et al. \cite{TIS} formulated a framework of multi-objective online 
problems by using that of multi-objective optimization problems \cite{E}. 
In this subsection, we present 
multi-objective {\it maximization\/} problems
(multi-objective minimization problems can be defined analogously). 

Let ${\cal P}_{k}=({\cal I},{\cal X}, h)$ be  a multi-objective 
optimization (maximization) problem, 
where ${\cal I}$ is a set of inputs, 
${\cal X}(I) \subseteq {\bf R}^{k}$ 
is a set of feasible solutions for each input $I \in {\cal I}$, and 
$h: {\cal I} \times {\cal X} \to {\bf R}^{k}$ is a function such that 
$h(I,\vec{x}) \in {\bf R}^{k}$ represents the objective of 
each solution $\vec{x} \in {\cal X}(I)$. For an input $I \in {\cal I}$, 
an algorithm 
$\msc{alg}_{k}$ for 
${\cal P}_{k}$ computes a feasible solution $\msc{alg}_{k}[I] \in {\cal X}(I)$. 
For an input $I \in {\cal I}$ 
and each feasible solution $\msc{alg}_{k}[I] \in {\cal X}(I)$, 
let $\msc{alg}_{k}(I)=h(I,\msc{alg}_{k}[I]) \in {\bf R}^{k}$ be the objective 
associate with $\msc{alg}_{k}[I]$. 
We say that a feasible 
solution $\vec{x}_{\rm max} \in {\cal X}(I)$ is {\it maximal\/} if there 
exists no feasible solution 
$\vec{x} \in {\cal X}(I)\setminus\{\vec{x}_{\rm max}\}$ 
such that $h(I,\vec{x}_{\rm max}) \preceq h(I,\vec{x})$ and say that an algorithm $\msc{opt}_{k}$ for ${\cal P}_{k}$ is {\it optimal\/} if 
for any input $I \in {\cal I}$, 
$\msc{opt}_{k}[I] \subseteq {\bf R}^{k}$ is the set of maximal 
solutions for the input $I \in {\cal I}$, i.e., 
$\msc{opt}_{k}[I] = \{\vec{x} \in {\cal X}(I): \vec{x} 
\mbox{ is a maximal solution for } I \in {\cal I}\}$. 
We use $\msc{opt}_{k}(\vec{x})\in {\bf R}^{k}$ to denote the 
objective associated with a solution $\vec{x} \in \msc{opt}_{k}[I]$. 

A multi-objective online problem can be defined in a way similar
to a single-objective online problem \cite{BE}. We regard a multi-objective 
online problem as a multi-objective optimization~problem  
in which the input is 
revealed bit by bit and an output must be produced~in~an~online~manner, i.e., 
after each new part of input is revealed, a decision affecting the output 
must be made. 
%
%======================================================================
\subsection{Competitive Analysis for Multi-Objective Online Problems}
\label{subsec-competitive}
%======================================================================
%
Tiedemann, et al. \cite{TIS} defined a notion of competitive analysis for 
multi-objective online problems. 
In this subsection, we introduce 
the notion of competitive analysis for multi-objective~online 
problems with respect to maximization problems (it is straightforward that 
the corresponding minimization problem can be defined analogously). 
\begin{definition}[\cite{TIS}] \label{df-c-competitive}
Let ${\cal P}_{k}=({\cal I},{\cal X},h)$ be a multi-objective optimization 
%$($maximization$)$ 
problem. 
For a vector $\vec{c}=(c_{1},\ldots,c_{k}) \in {\bf R}^{k}$, 
we say that a multi-objective online algorithm $\msc{alg}_{k}$ for ${\cal P}_{k}$ 
is {\sf $\vec{c}$-competitive} if 
for every input sequence $I\in {\cal I}$, there exists 
a maximal solution $\vec{x} \in \msc{opt}_{k}[I]$ such that 
\[
\bigwedge_{i \in [1,k]} \left[\msc{opt}_{k}(\vec{x})_{i} \leq 
c_{i} \cdot  \msc{alg}_{k}(I)_{i} + \alpha_{i} \right], 
\]
%
%for each $i \in [1,k]$, 
where $\vec{\alpha}=(\alpha_{1},\ldots,\alpha_{k}) \in {\bf R}^{k}$ is a 
constant vector independent of input sequences $I \in {\cal I}$. 
\end{definition}

It should be noted that for multi-objective online algorithms, the notion of 
$\vec{c}$-competitive~is defined by a vector 
$\vec{c}=(c_{1},\ldots,c_{k}) \in {\bf R}^{k}$, while for 
single-objective online algorithms,~the~notion of $c$-competitive is defined by 
a single scalar $c \geq 1$. 

\begin{definition}[\cite{TIS}] \label{df-s-c-competitive}
Let ${\cal P}_{k}=({\cal I},{\cal X},h)$ be a multi-objective optimization 
%$($maximization$)$ 
problem. 
For a vector $\vec{c}=(c_{1},\ldots,c_{k}) \in {\bf R}^{k}$, 
we say that a multi-objective online algorithm $\msc{alg}_{k}$ for ${\cal P}_{k}$ 
is {\sf strongly $\vec{c}$-competitive} if 
for every input sequence $I \in {\cal I}$ and every maximal 
solution $\vec{x} \in \msc{opt}_{k}[I]$, 
\[
\bigwedge_{i \in [1,k]} \left[
\msc{opt}_{k}(\vec{x})_{i} \leq c_{i} \cdot  \msc{alg}_{k}(I)_{i} + \alpha_{i}
\right],
\]
%
%holds for each $i \in [1,k]$, 
where 
$\vec{\alpha}=(\alpha_{1},\ldots,\alpha_{k}) \in {\bf R}^{k}$ is a constant 
vector independent of input sequences $I \in {\cal I}$. 
\end{definition}

Let $f: {\bf R}^{k} \to {\bf R}$ be a monotone function. For a 
multi-objective online algorithm $\msc{alg}_{k}$~for~${\cal P}_{k}$, 
the {\it competitive ratio\/} of $\msc{alg}_{k}$ 
with respect to  $f$ is the infimum of $f(\vec{c})$ over all possible 
vectors $\vec{c}=(c_{1},\ldots,c_{k}) \in {\bf R}^{k}$   
such that $\msc{alg}_{k}$ is $\vec{c}$-competitive. Let 
${\cal C}[\msc{alg}_{k}]$ be the set of all possible~vectors 
$\vec{c} = (c_{1},\ldots,c_{k}) \in {\bf R}^{k}$ such that 
$\msc{alg}_{k}$ is $\vec{c}$-competitive and 
${\cal C}_{s}[\msc{alg}_{k}]$ be the set of all possible vectors 
$\vec{c} = (c_{1},\ldots,c_{K}) \in {\bf R}^{k}$ such that 
$\msc{alg}_{k}$ is strongly  $\vec{c}$-competitive, i.e., 
\begin{eqnarray*}
{\cal C}[\msc{alg}_{k}] & = & \{\vec{c} \in {\bf R}^{k}: \msc{alg}_{k} 
\mbox{ is $\vec{c}$-competitive}\};\\
{\cal C}_{s}[\msc{alg}_{k}] & = & \{\vec{c} \in {\bf R}^{k}: \msc{alg}_{k} 
\mbox{ is strongly $\vec{c}$-competitive}\}. 
\end{eqnarray*}
\begin{definition}[\cite{TIS}] \label{df-ratio}
Let $f: {\bf R}^{k} \to {\bf R}$ be a monotone function and 
$\msc{alg}_{k}$ be an online algorithm for 
a multi-objective optimization $($maximization$)$ problem 
${\cal P}_{k}$. 
The {\sf competitive ratio} of the algorithm $\msc{alg}_{k}$ 
with respect to $f$ is 
\[
{\cal R}^{f}(\msc{alg}_{k}) = 
\inf_{\vec{c} \in {\cal C}[{\tiny {\rm ALG}}_{k}]} f(\vec{c}),
\]
and the {\sf strong competitive ratio} of 
the algorithm $\msc{alg}_{k}$ with respect to $f$ is 
\[
{\cal R}_{s}^{f}(\msc{alg}_{k}) 
= \inf_{\vec{c} \in {\cal C}_{s}[{\tiny {\rm ALG}}_{k}]} f(\vec{c}). 
\]
\end{definition} 

Natural examples of a monotone function $f: {\bf R}^{k} \to {\bf R}$ are 
given by Tiedemann, et al. \cite{TIS}: 
\begin{eqnarray*}
f_{1}(c_{1},\ldots,c_{k}) & = & \max \left(c_{1},\ldots,c_{k}\right);\\
f_{2}(c_{1},\ldots,c_{k}) & = & \frac{1}{k}(c_{1}+\cdots+c_{k});\\
f_{3}(c_{1},\ldots,c_{k}) & = & 
\left(c_{1} \times \cdots \times c_{k} \right)^{1/k}. 
\end{eqnarray*}
%
%We also know 
Another example of a monotone function is 
%$f_{4}:{\bf R}^{k}\to {\bf R}$, i.e., 
$f_{4}(c_{1},\ldots,c_{k})=\min(c_{1},\ldots,c_{k})$. 
We refer~to  the competitive ratio of an algorithm $\msc{alg}_{k}$ 
with respect to functions 
$f_{1}$,~$f_{2}$,~$f_{3}$,~and~$f_{4}$~as~the {\it worst component\/} 
competitive ratio, the {\it arithmetic mean component\/} competitive ratio, 
the~{\it geometric mean component\/} competitive ratio, and 
the {\it best component\/} competitive ratio, respectively. 
Note that all of the monotone functions $f_{1}$, $f_{2}$, $f_{3}$, and $f_{4}$ 
are continuous on ${\bf R}^{k}$~for~any~$k \geq 1$. 
%
%=========================================================
\subsection{Multi-Objective Time Series Search Problem}
\label{subsec-time-series}
%=========================================================
%
A single-objective time series search problem is initially investigated 
by El-Yaniv, et al. \cite{Eetal}~and~it is defined as follows: An online 
player $\msc{alg}$ is searching for the maximum price in a sequence~of prices. 
At the beginning of each time period $t \in [1,T]$, a price $p_{t}$ is revealed 
to the online~player $\msc{alg}$ and it must decide 
whether to accept or reject the price $p_{t}$. 
If the online player 
$\msc{alg}$~accepts the price $p_{t}$, then 
the game ends and the return for 
$\msc{alg}$ is $p_{t}$. We assume that prices are chosen from 
the interval $\msc{itv}=[m,M]$, where 
$0<m \leq M$, and that $m$ and $M$ are known to the online player 
$\msc{alg}$\footnote{ It is possible to show that 
if only the fluctuation ratio 
$\phi=M/m$ is known (but not $m$ or $M$) 
to the online player {\sc alg}, then no better competitive  
ratio than the trivial one of $\phi$ is achievable.}. 
If the online player $\msc{alg}$ rejects the price $p_{t}$ 
for every $t \in [1,T]$, 
then the return~for 
$\msc{alg}$ is defined to be $m$. 
A multi-objective time series search 
problem \cite{TIS}~can~be~defined~by~a~na\-tural extension of the 
single-objective time series search problem. 

In a multi-objective %($k$-objective) 
time series search problem, a price vector 
$\vec{p}_{t}=(p_{t}^{1},\ldots,p_{t}^{k}) \in {\bf R}^{k}$~is~revealed 
to the online player $\msc{alg}_{k}$ 
at the beginning of each time period $t \in [1,T]$,~and~the~online 
player $\msc{alg}_{k}$ must decide whether
to accept or reject the price vector $\vec{p}_{t}$. 
If the online player~$\msc{alg}_{k}$ accepts the price vector $\vec{p}_{t}$, 
then the game ends and the return for 
$\msc{alg}_{k}$ is 
$\vec{p}_{t}$. 
As in the case~of~a single-objective time series search problem, 
assume that prices $p_{t}^{i}$ are chosen from the interval 
$\msc{itv}_{i}=[m_{i},M_{i}]$ with $0 < m_{i} \leq M_{i}$ for each $i \in [1,k]$, 
and that 
the online~player~$\msc{alg}_{k}$~knows~$m_{i}$ and $M_{i}$ for each $i \in [1,k]$. 
If the online player $\msc{alg}_{k}$ rejects the price vector 
$\vec{p}_{t}$ for every $t \in [1,T]$, then 
the return for of the online player $\msc{alg}_{k}$ 
is defined to be the {\it minimum\/} price vector 
$\vec{p}_{\rm min}=(m_{1},\ldots,m_{k})$. 
Without loss of generality, we assume that 
$M_{1}/m_{1}\geq \cdots \geq M_{k}/m_{k}$. 
%
%====================================================
\section{Observations on the Competitive Analysis} 
\label{sec-observation}
%====================================================
%
For the multi-objective ($k$-objective) time series search problem, 
it is natural to regard that~$m_{i}$ and $M_{i}$ 
are part of the problem (not part of input sequences) 
for each $i \in [1,k]$. By setting~$\alpha_{i}=M_{i}$ 
(as a constant independent of input sequences) 
for each $i \in [1,k]$, 
we can take $c_{1}=\cdots = c_{k}=0$ 
in Definitions \ref{df-c-competitive} and \ref{df-s-c-competitive}. 
This implies that any algorithm {\sc alg} for the multi-objective 
($k$-objective) time 
series search problem is $(0,\ldots,0)$-competitive, i.e., 
for any monotone function $f:{\bf R}^{k} \to {\bf R}$, 
the competitive ratio of the algorithm {\sc alg} is $f(0,\ldots,0)$. 
Thus in Definitions \ref{df-c-competitive} and \ref{df-s-c-competitive}, 
we fix $\alpha_{i}=0$ for each $i \in [1,k]$. 

For simplicity, assume that $k=2$ and $I_{1}=I_{2}=[m,M]$, where $0 < m < M$. 
Consider~a~simple algorithm $\msc{alg}_{2}$ that accepts the 
first price vector for any input sequence and~observe~how~the 
competitive analysis for the algorithm $\msc{alg}_{2}$ 
works in the following examples: 
\begin{example} \label{example-1}
{\rm 
Let ${\cal I}_{1}=\{s_{1},s_{2}\}$ be the set of input sequences. 
In the input sequence $s_{1}$, price vectors $\vec{p}_{1}=(m,M)$,  
$\vec{p}_{2}=(M,m)$, and $\vec{p}_{3}=(m,m)$ are revealed 
to the algorithm $\msc{alg}_{2}$~at~$t=1$, $t=2$, and $t=3$, 
respectively, and in the input sequence $s_{2}$, 
price~vectors~$\vec{q}_{1}=(M,m)$,~$\vec{q}_{2}=(m,m)$, and 
$\vec{q}_{3}=(m,M)$ 
are revealed to the algorithm $\msc{alg}_{2}$ 
at $t=1$,~$t=2$,~and~$t=3$,~respec\-tively. 
%
%Note that 
For the input sequence $s_{1}$, 
the algorithm $\msc{alg}_{2}$ accepts 
$\vec{p}_{1}=(m,M)$ which is maximal~in $s_{1}$ and for the 
input sequence $s_{2}$, the algorithm $\msc{alg}_{2}$ accepts 
$\vec{p}_{2}=(M,m)$ which~is~also~maximal in $s_{2}$. 
From Definition \ref{df-s-c-competitive}, we have that 
the algorithm $\msc{alg}_{2}$ is strongly 
$(\frac{M}{m},\frac{M}{m})$-competitive. 
}
\end{example}
\begin{example} \label{example-2}
{\rm
Let ${\cal I}_{2}=\{\sigma\}$ be the set of input sequences. 
In the input sequence $\sigma$,~price~vectors $\vec{r}_{1}=(m,m)$, 
$\vec{r}_{2}=(m,M)$, and $\vec{r}_{3}=(M,m)$ are revealed 
at $t=1$, $t=2$, and $t=3$~to~the~algorithm $\msc{alg}_{2}$, respectively. 
%Notice that 
The algorithm $\msc{alg}_{2}$ 
accepts $\vec{r}_{1}=(m,m)$ which is not maximal~in~$\sigma$. 
From Definition \ref{df-s-c-competitive}, 
we have that the algorithm $\msc{alg}_{2}$ is strongly 
$(\frac{M}{m},\frac{M}{m})$-competitive. 
}
\end{example}

In Example \ref{example-1}, 
the algorithm $\msc{alg}_{2}$ accepts price vectors which 
is maximal in the input sequences $s_{1}$ and $s_{2}$,  however, in Example 
\ref{example-2}, the algorithm $\msc{alg}_{2}$ accepts 
a price vector which is not maximal in the input sequence $\sigma$. 
Thus it follows that 
for any monotone function~$f:{\bf R}^{2} \to {\bf R}$,  
the strong competitive ratio of the algorithm $\msc{alg}_{2}$ is 
$f(M/m,M/m)$ for both Examples \ref{example-1} and \ref{example-2}, which 
does not necessarily capture the efficiency of 
online algorithms. To derive a more realistic framework, we 
need to modify the definition of competitive ratio. 

Let $\msc{alg}_{k}$ be an online algorithm for a multi-objective 
optimization (maximization) problem ${\cal P}_{k}$. 
We use ${\cal CR}^{f}(\msc{alg}_{k};I)$ to denote the competitive ratio 
of the algorithm $\msc{alg}_{k}$ for~an~input~sequence 
$I \in {\cal I}$ with respect to a monotone 
function $f : {\bf R}^{k} \to {\bf R}$, i.e., 
\[
{\cal CR}^{f}(\msc{alg}_{k};I)=\sup_{\vec{x} \in \msc{opt}_{k}[I]} 
f\left(
\frac{\msc{opt}_{k}(\vec{x})_{1}}{\msc{alg}_{k}(I)_{1}}, \ldots,
\frac{\msc{opt}_{k}(\vec{x})_{k}}{\msc{alg}_{k}(I)_{k}} \right). 
\]
\begin{definition} \label{df-mod-ratio}
Let $\msc{alg}_{k}$ be  a multi-objective online algorithm for ${\cal P}_{k}$. 
The {\sf competitive ratio} of the algorithm $\msc{alg}_{k}$ 
with respect to a monotone function $f: {\bf R}^{k} \to {\bf R}$ is 
\[
{\cal CR}^{f}(\msc{alg}_{k}) = \sup_{I \in {\cal I}} 
{\cal CR}^{f}(\msc{alg}_{k};I).
\]
\end{definition}

It is easy to see that for the case that all of $\msc{itv}_{1}=[m_{1},M_{1}], 
\ldots,\msc{itv}_{k}=[m_{k},M_{k}]$~are~{\it real\/}~intervals, 
%In fact, 
all of the analyses on the 
competitive ratio by Tiedemann, et al. 
\cite{TIS} hold under Definition \ref{df-mod-ratio}. 
In the rest of the paper, we analyze the algorithms under Definition 
\ref{df-mod-ratio}.  
%
%==============================================
\section{Online Algorithm: Balanced Price Policy} 
\label{sec-bpp}
%==============================================
%
As mentioned in Section \ref{sec-introduction}, 
Tiedemann, et al. \cite{TIS} presented some
online algorithms for the multi-objective ($k$-objective) 
time series search problem and analyzed 
the competitive ratio~of~those algorithms  
with respect to the monotone functions 
$f_{1}$, $f_{2}$, and $f_{3}$. The competitive analysis given in \cite{TIS} 
heavily depends on the fact that the monotone functions 
$f_{1}$, $f_{2}$, and $f_{3}$ are continuous and the assumption that 
all of $\msc{itv}_{1}=[m_{1},M_{1}],\ldots,\msc{itv}_{k}=[m_{k},M_{k}]$ are real 
intervals. 

In this section, we present a simple online algorithm Balanced Price Policy 
$\msc{bpp}_{k}$ (in Figure \ref{fig-bpp}) for the multi-objective ($k$-objective) time 
series search problem with respect to an arbitrary monotone 
function $f: {\bf R}^{k} \to {\bf R}$. 
\begin{figure*}[htb]
\begin{center}
\fbox{
\begin{minipage}{11.25cm}\smallskip
\begin{tabular}{l}
{\bf for} $t=1,2,\ldots, T$ {\bf do}\\
   $|$\\[-0.25cm]
   $|$  Accept $\vec{p}_{t} = (p_{t}^{1},\ldots,p_{t}^{k})$ if 
$f(\frac{M_{1}}{p_{t}^{1}},\ldots,\frac{M_{k}}{p_{t}^{k}}) \leq 
f(\frac{p_{t}^{1}}{m_{1}},\ldots,\frac{p_{t}^{k}}{m_{k}})$. \\[-0.25cm]
   $|$\\
{\bf end}
\end{tabular}
\end{minipage}
}\bigskip\\
\caption{Balanced Price Policy $\msc{bpp}_{k}$} \label{fig-bpp}
\end{center}
\end{figure*}
%
%===================================================
\subsection{General Case} \label{subsec-general}
%===================================================
%
In this subsection, we do not assume that all of 
$\msc{itv}_{1}=[m_{1},M_{1}],\ldots,\msc{itv}_{k}=[m_{k},M_{k}]$ 
are~real~intervals (in fact, $\msc{itv}_{i}=[m_{i},M_{i}]$ is allowed 
to be an integral interval) 
and we deal with~any~monotone (not necessarily continuous) 
function $f:{\bf R}^{k}\to {\bf R}$. 

\begin{theorem} \label{thm-general}
Let $\msc{alg}_{k}$ be an arbitrary online algorithm for 
the multi-objective $(k$-objective$)$ time series search problem. Then 
${\cal CR}^{f}(\msc{bpp}_{k})\leq 
{\cal CR}^{f}(\msc{alg}_{k})$ for any monotone 
$($not necessarily continuous$)$ function 
$f : {\bf R}^{k}\to {\bf R}$ and any integer $k \geq 1$. 
\end{theorem}
{\bf Proof:} We use  
$I=(\vec{p}_{1},\ldots,\vec{p}_{T})$ to denote an arbitrary input sequence, 
where $\vec{p}_{t} = (p_{t}^{1},\ldots,p_{t}^{k}) \in 
\msc{itv}_{1}\times \cdots \times \msc{itv}_{k}$ 
for each $t \in [1,T]$. Let 
${\cal I}$ be the set of input sequences. 
Define ${\cal I}_{\rm acc} \subseteq {\cal I}$ to be the  
set of input sequences accepted by the algorithm $\msc{bpp}_{k}$ and 
${\cal I}_{\rm rej} \subseteq {\cal I}$ to be the  
set of input sequences rejected by the algorithm $\msc{bpp}_{k}$, i.e., 
\begin{eqnarray*}
{\cal I}_{\rm acc} & = & \left \{
( \vec{p}_{1},\ldots,\vec{p}_{T}) \in {\cal I}: 
\bigvee_{t \in [1,T]} \left[
f\left(\frac{M_{1}}{p_{t}^{1}},\ldots,
\frac{M_{k}}{p_{t}^{k}}\right) \leq 
f\left(\frac{p_{t}^{1}}{m_{1}},\ldots,\frac{p_{t}^{k}}{m_{k}}\right) \right]\right\};\\
{\cal I}_{\rm rej} & = & \left \{
( \vec{p}_{1},\ldots,\vec{p}_{T}) \in {\cal I}: 
\bigwedge_{t \in [1,T]} \left[
f\left(\frac{M_{1}}{p_{t}^{1}},\ldots,
\frac{M_{k}}{p_{t}^{k}}\right) > 
f\left(\frac{p_{t}^{1}}{m_{1}},\ldots,\frac{p_{t}^{k}}{m_{k}}\right) \right]\right\}. 
\end{eqnarray*}
Let $\msc{alg}_{k}$ be an arbitrary online algorithm for 
the multi-objective time series search problem. 

For each $I = ( \vec{p}_{1},\ldots,\vec{p}_{T}) \in {\cal I}_{\rm acc}$, 
the algorithm $\msc{bpp}_{k}$ halts at the earliest time 
$t[I] \in [1,T]$ to accept a price vector
$\vec{p}_{t[I]}=(p_{t[I]}^{1},\ldots,p_{t[I]}^{k})$ such that 
\[
f\left(\frac{M_{1}}{p_{t[I]}^{1}},\ldots,\frac{M_{k}}{p_{t[I]}^{k}}\right) \leq 
f\left(\frac{p_{t[I]}^{1}}{m_{1}},\ldots,\frac{p_{t[I]}^{k}}{m_{k}}\right), 
\]
and let $I^{*}=(\vec{p}_{t[I]},\vec{p}_{\rm max})$, where $\vec{p}_{\rm max}=
(M_{1},\ldots,M_{k})$. For each $I =(\vec{p}_{1},\ldots,\vec{p}_{T}) 
\in {\cal I}_{\rm acc}$,~it~is~immediate that 
$I^{*}=(\vec{p}_{t[I]},\vec{p}_{\rm max}) \in {\cal I}_{\rm acc}$ and 
\begin{eqnarray}
{\cal CR}^{f}(\msc{bpp}_{k};I) & = &
\max_{\vec{x} \in \msc{opt}_{k}[I]} 
f\left(
\frac{\msc{opt}_{k}(\vec{x})_{1}}{\msc{bpp}_{k}(I)_{1}}, \ldots,
\frac{\msc{opt}_{k}(\vec{x})_{k}}{\msc{bpp}_{k}(I)_{k}} \right)\nonumber\\
& = & 
\max_{\vec{x} \in \msc{opt}_{k}[I]} 
f\left(
\frac{\msc{opt}_{k}(\vec{x})_{1}}{p_{t[I]}^{1}}, \ldots,
\frac{\msc{opt}_{k}(\vec{x})_{k}}{p_{t[I]}^{k}} \right)\nonumber\\
& \leq & 
f\left(\frac{M_{1}}{p_{t[I]}^{1}}, \ldots,
\frac{M_{k}}{p_{t[I]}^{k}} \right)
= {\cal CR}^{f}(\msc{bpp}_{k};I^{*}), \label{eq-acc-bound}
\end{eqnarray}
where the inequality follows from the assumption that %the function 
$f: {\bf R}^{k}\to {\bf R}$ is monotone. 
Let~${\cal I}^{*}_{\rm acc}=
\{ I^{*} = (\vec{p}_{t[I]},\vec{p}_{\rm max}): 
I \in {\cal I}_{\rm acc}\}$. Note that 
${\cal I}_{\rm acc}^{*} \subseteq {\cal I}_{\rm acc}$. 
For each $J^{*} =(\vec{p},\vec{p}_{\rm max}) 
\in {\cal I}^{*}_{\rm acc}$,~define~$J'$~according to 
how the algorithm $\msc{alg}_{k}$ works 
on receiving the price vector $\vec{p}=(p^{1},\ldots,p^{k})$. 
For~the case that the algorithm $\msc{alg}_{k}$ accepts the price vector 
$\vec{p}$, 
let $J'=(\vec{p},\vec{p}_{\rm max})$ and we have that 
\[
{\cal CR}^{f}(\msc{bpp}_{k};J^{*}) = 
f \left(\frac{M_{1}}{p^{1}},\ldots,\frac{M_{k}}{p^{k}}\right)
= {\cal CR}^{f}(\msc{alg}_{k};J'). 
\]
For the case that the algorithm 
$\msc{alg}_{k}$ rejects the price vector 
$\vec{p}$, 
let $J'=(\vec{p})$ and we have that 
\[
{\cal CR}^{f}(\msc{bpp}_{k};J^{*}) = 
f \left(\frac{M_{1}}{p^{1}},\ldots,\frac{M_{k}}{p^{k}}\right)
\leq f\left(\frac{p^{1}}{m_{1}},\ldots,\frac{p^{k}}{m_{k}}\right)
= {\cal CR}^{f}(\msc{alg}_{k};J'), 
\]
where the inequality is due to the assumption that 
$J^{*}=(\vec{p},\vec{p}_{\rm max}) \in {\cal I}_{\rm acc}$, i.e., 
the algorithm $\msc{bpp}_{k}$ accepts %the price vector 
$\vec{p}=(p^{1},\ldots,p^{k})$ by %due to 
the condition that  $f(M_{1}/p^{1},\ldots,M_{k}/p^{k}) \leq 
f(p^{1}/m_{1},\ldots,p^{k}/m_{k})$. 
Thus for each $I \in {\cal I}_{\rm acc}$, 
there exists a price vector $I' \in {\cal I}$ such that 
\begin{equation}
{\cal CR}^{f}(\msc{bpp}_{k};I)\leq {\cal CR}^{f}(\msc{alg}_{k};I'). 
\label{eq-acc-alg}
\end{equation}

For each $I =(\vec{p}_{1},\ldots,\vec{p}_{T}) \in {\cal I}_{\rm rej}$, 
the algorithm $\msc{bpp}_{k}$ rejects a price vector 
$\vec{p}_{t}$~for~every~$t \in [1,T]$, i.e., 
$f(M_{1}/p_{t}^{1},\ldots,M_{k}/p_{t}^{k}) > 
f(p_{t}^{1}/m_{1}, \ldots,p_{t}^{k}/m_{k})$ for every $t \in [1,T]$, 
and settles in the 
minimum price vector $\vec{p}_{\rm min} = ( m_{1},\ldots,m_{k})$.  
At time $\tau[I] \in [1,T]$, however, 
the optimal offline algorithm 
$\msc{opt}_{k}$ can accept a price vector 
$\vec{p}_{\tau[I]}=(p_{\tau[I]}^{1},\ldots,p_{\tau[I]}^{k})$ such that 
%$f(p_{\tau[I]}^{1}]m_{1},\ldots,p_{\tau[I]}^{k}/m_{k})=
%\max_{t \in [1,T]} 
%f(p_{t}^{1}/m_{1},\ldots,p_{t}^{k}/m_{k})$, 
%
\[
f\left(\frac{p_{\tau[I]}^{1}}{m_{1}},\ldots,\frac{p_{\tau[I]}^{k}}{m_{k}}\right)=
\max_{t \in [1,T]} 
f\left(\frac{p_{t}^{1}}{m_{1}},\ldots,\frac{p_{t}^{k}}{m_{k}}\right), 
\]
and let $I^{*}=(\vec{p}_{\tau[I]})$. 
For each $I =(\vec{p}_{1},\ldots,\vec{p}_{T}) 
\in {\cal I}_{\rm rej}$, it is immediate that 
$I^{*}=(\vec{p}_{\tau[I]}) \in {\cal I}_{\rm rej}$ and 
\begin{eqnarray}
{\cal CR}^{f}(\msc{bpp}_{k};I) & = & 
\max_{\vec{x} \in \msc{opt}_{k}[I]} 
f\left(
\frac{\msc{opt}_{k}(\vec{x})_{1}}{\msc{bpp}_{k}(I)_{1}}, \ldots,
\frac{\msc{opt}_{k}(\vec{x})_{k}}{\msc{bpp}_{k}(I)_{k}} \right)\nonumber\\
& = & 
\max_{\vec{x} \in \msc{opt}_{k}[I]}
f\left(
\frac{\msc{opt}_{k}(\vec{x})_{1}}{m_{1}}, \ldots,
\frac{\msc{opt}_{k}(\vec{x})_{k}}{m_{k}} \right)\nonumber\\
& =& 
f\left(\frac{p_{\tau[I]}^{1}}{m_{1}}, \ldots,
\frac{p_{\tau[I]}^{k}}{m_{k}} \right)
= {\cal CR}^{f}(\msc{bpp}_{k};I^{*}). \label{eq-rej-bound}
\end{eqnarray}
Let ${\cal I}^{*}_{\rm rej}=
\{ I^{*} = (\vec{p}_{\tau[I]}): I \in {\cal I}_{\rm rej}\}$. Note that 
${\cal I}_{\rm rej}^{*} \subseteq {\cal I}_{\rm rej}$.  
For each $J^{*} =(\vec{p}) 
\in {\cal I}^{*}_{\rm rej}$, 
define~$J'$~according to how the algorithm $\msc{alg}_{k}$ works 
on receiving the price vector $\vec{p}=(p^{1},\ldots,p^{k})$. 
For~the case that the algorithm $\msc{alg}_{k}$ accepts the price vector 
$\vec{p}$, 
let $J'=(\vec{p},\vec{p}_{\rm max})$ and we have that 
\[
{\cal CR}^{f}(\msc{bpp}_{k};J^{*}) = 
f \left(\frac{p^{1}}{m_{1}},\ldots,\frac{p^{k}}{m_{k}}\right)
< f \left(\frac{M_{1}}{p^{1}},\ldots,\frac{M_{k}}{p^{k}}\right)
= {\cal CR}^{f}(\msc{alg}_{k};J'), 
\]
where the inequality is due to the assumption that 
$J^{*}=(\vec{p}) \in {\cal I}_{\rm rej}$, i.e., 
the algorithm~$\msc{bpp}_{k}$ rejects %the price vector 
$\vec{p}=(p^{1},\ldots,p^{k})$ 
by the condition that $f(M_{1}/p^{1},\ldots,M_{k}/p^{k}) > f(p^{1}/m_{1},
\ldots,p^{k}/m_{k})$. 
For the case that the algorithm $\msc{alg}_{k}$ rejects the price vector 
$\vec{p}$, let $J'=(\vec{p})$ and we have that 
\[
{\cal CR}^{f}(\msc{bpp}_{k};J^{*}) = 
f \left(\frac{p^{1}}{m_{1}},\ldots,\frac{p^{k}}{m_{k}}\right)
= {\cal CR}^{f}(\msc{alg}_{k};J'). 
\]
Thus for each $I \in {\cal I}_{\rm rej}$, 
there exists a price vector $I' \in {\cal I}$ such that 
\begin{equation}
{\cal CR}^{f}(\msc{bpp}_{k};I)\leq {\cal CR}^{f}(\msc{alg}_{k};I'). 
\label{eq-rej-alg}
\end{equation}
Then from Definition \ref{df-mod-ratio}, 
it follows that 
\begin{eqnarray*}
{\cal CR}^{f}(\msc{bpp}_{k}) & = & 
\sup_{I \in {\cal I}} {\cal CR}^{f}(\msc{bpp}_{k};I)\\
& = & \max\left\{
\sup_{I \in {\cal I}_{\rm acc}} {\cal CR}^{f}(\msc{bpp}_{k};I), 
\sup_{I \in {\cal I}_{\rm rej}} 
{\cal CR}^{f}(\msc{bpp}_{k};I)\right\}\\ 
& \leq & \max\left\{
\sup_{I' \in {\cal I}} {\cal CR}^{f}(\msc{alg}_{k};I'),
\sup_{I' \in {\cal I}} {\cal CR}^{f}(\msc{alg}_{k};I')\right\}\\
& = & \sup_{I' \in {\cal I}} {\cal CR}^{f}(\msc{alg}_{k};I') = {\cal CR}^{f}
(\msc{alg}_{k}), 
\end{eqnarray*}
where the inequality follows from 
Equations (\ref{eq-acc-alg}) and (\ref{eq-rej-alg}). \BQED
%
%=========================================================
\subsection{Special Case: %Real Intervals and 
Monotone Continuous Functions} \label{subsec-special}
%=========================================================
%
In this subsection, we assume that all of $\msc{itv}_{1}=[m_{1},M_{1}],
\ldots,\msc{itv}_{k}=[m_{k},M_{k}]$ are real intervals and deal with 
only monotone continuous functions $f : {\bf R}^{k}\to {\bf R}$. 

Let $z_{f}^{k} = \sup_{(x_{1},\ldots,x_{k}) \in  {\cal S}_{f}^{k}} 
f(M_{1}/x_{1},\ldots,M_{k}/x_{k})$, where 
\[
{\cal S}_{f}^{k} =  \left\{(x_{1},\ldots,x_{k})\in 
\msc{itv}_{1} \times \cdots \times \msc{itv}_{k}: 
f\left(\frac{M_{1}}{x_{1}}, \ldots, 
\frac{M_{k}}{x_{k}}\right)=f\left(\frac{x_{1}}{m_{1}}, \ldots, 
\frac{x_{k}}{m_{k}}\right)\right\}. 
\]
By setting $x_{i}=\sqrt{m_{i}M_{i}}\in I_{i}=[m_{i},M_{i}]$ for each $i \in [1,k]$, 
we have that 
\begin{eqnarray*}
f\left(\frac{M_{1}}{x_{1}},\ldots,\frac{M_{k}}{x_{k}}\right) & = & 
f\left(\frac{M_{1}}{\sqrt{m_{1}M_{1}}},\ldots,\frac{M_{k}}{\sqrt{m_{k}M_{k}}}\right)  =
f\left(\sqrt{\frac{M_{1}}{m_{1}}},\ldots,\sqrt{\frac{M_{k}}{m_{k}}}\right);\\
f\left(\frac{x_{1}}{m_{1}},\ldots,\frac{x_{k}}{m_{k}}\right) & = & 
f\left(\frac{\sqrt{m_{1}M_{1}}}{m_{1}},\ldots,\frac{\sqrt{m_{k}M_{k}}}{m_{k}}\right)  =
f\left(\sqrt{\frac{M_{1}}{m_{1}}},\ldots,\sqrt{\frac{M_{k}}{m_{k}}}\right).
\end{eqnarray*}
Thus for any monotone continuous 
function $f$, it follows that 
$(\sqrt{m_{1}M_{1}},\ldots,\sqrt{m_{k}M_{k}}) \in {\cal S}_{f}^{k}$. So  
we have that ${\cal S}_{f}^{k} \neq \emptyset$ and 
$z_{f}^{k} = \sup_{(x_{1},\ldots,x_{k}) \in  {\cal S}_{f}^{k}} 
f(M_{1}/x_{1},\ldots,M_{k}/x_{k})$ is well-defined. 

In this subsection, we show that the the exact value of the competitive ratio 
of the algorithm $\msc{bpp}_{k}$ is $z_{f}^{k}$ 
for any monotone {\it continuous\/} function $f: {\bf R}^{k} \to {\bf R}$ and 
any integer $k \geq 1$~(Corollary \ref{cor-unified}).  
More precisely, we show that 
${\cal CR}^{f}(\msc{bpp}_{k}) \leq z_{f}^{k}$ (Theorem 
\ref{thm-unified-upper}) 
and that 
${\cal CR}^{f}(\msc{alg}_{k}) \geq z_{f}^{k}$~for any algorithm $\msc{alg}_{k}$
(Theorem \ref{thm-unified-lower}). From Theorem \ref{thm-general} and 
Corollary \ref{cor-unified},~it~follows~that~$z_{f}^{k}$~is~the 
best possible value of the competitive ratio for the multi-objective %($k$-objective) 
time series search problem. 
\begin{theorem} \label{thm-unified-upper}
If all of $\msc{itv}_{1}=[m_{1},M_{1}],\ldots,\msc{itv}_{k}=[m_{k},M_{k}]$ 
are real intervals, then 
for any monotone continuous 
function $f: {\bf R}^{k}\to {\bf R}$ and any integer $k \geq 1$, 
${\cal CR}^{f}(\msc{bpp}_{k}) \leq z_{f}^{k}$. 
\end{theorem}
{\bf Proof:} Let 
$I=(\vec{p}_{1},\ldots,\vec{p}_{T})$ to be an arbitrary input sequence, 
where $\vec{p}_{t} = (p_{t}^{1},\ldots,p_{t}^{k}) \in {\bf R}^{k}$~for 
each $t \in [1,T]$, and 
${\cal I}$ be the set of input sequences. 
As in the proof of Theorem \ref{thm-general}, we consider the set 
${\cal I}_{\rm acc} \subseteq {\cal I}$ of input sequences accepted by 
the algorithm $\msc{bpp}_{k}$ and the set 
${\cal I}_{\rm rej} \subseteq {\cal I}$~of~input sequences rejected by 
the algorithm $\msc{bpp}_{k}$. 

For each $I = ( \vec{p}_{1},\ldots,\vec{p}_{T}) \in {\cal I}_{\rm acc}$, 
the algorithm $\msc{bpp}_{k}$ halts at the earliest time 
$t[I] \in [1,T]$~to accept 
$\vec{p}_{t[I]}=(p_{t[I]}^{1},\ldots,p_{t[I]}^{k})$ such that 
$f(M_{1}/p_{t[I]}^{1},\ldots,M_{k}/p_{t[I]}^{k}) \leq 
f(p_{t[I]}^{1}/m_{1},\ldots,p_{t[I]}^{k}/m_{k})$. Thus from 
Equation (\ref{eq-acc-bound}), we have that 
\[ 
{\cal CR}^{f}(\msc{bpp}_{k};I) \leq 
f\left(\frac{M_{1}}{p_{t[I]}^{1}}, \ldots,
\frac{M_{k}}{p_{t[I]}^{k}} \right). 
\]
To show that 
$f(M_{1}/p_{t[I]}^{1},\ldots,M_{k}/p_{t[I]}^{k}) \leq z_{f}^{k}$, 
we consider the following cases: 
\begin{namelist}{~~~~~(2)}
\item[(1)] $f(M_{1}/p_{t[I]}^{1},\ldots,M_{k}/p_{t[I]}^{k}) = 
f(p_{t[I]}^{1}/m_{1},\ldots,p_{t[I]}^{k}/m_{k})$; 
\item[(2)] $f(M_{1}/p_{t[I]}^{1},\ldots,M_{k}/p_{t[I]}^{k}) <
f(p_{t[I]}^{1}/m_{1},\ldots,p_{t[I]}^{k}/m_{k})$. 
\end{namelist}
For the case (1), it is immediate that $\vec{p}_{t[I]} \in {\cal S}_{f}^{k}$ and 
$f(M_{1}/p_{t[I]}^{1},\ldots, M_{k}/p_{t[I]}^{k}) \leq  z_{f}^{k}$~by~definition. 
For the case (2), let 
${\cal J} = \{j \in [1,k]: M_{j}/p_{t[I]}^{j} \leq p_{t[I]}^{j}/m_{j}\}$. 
We claim~that~${\cal J} \neq \emptyset$\footnote{~By contradiction. 
If ${\cal J} = \emptyset$, 
then $M_{i}/p_{t[I]}^{i} > p_{t[I]}^{i}/m_{i}$ for each $i \in [1,k]$. 
Since the function $f: {\bf R}^{k}\to {\bf R}$ is monotone, we have that 
$f(M_{1}/p_{t[I]}^{1},\ldots,M_{k}/p_{t[I]}^{k}) \geq  
f(p_{t[I]}^{1}/m_{1},\ldots,p_{t[I]}^{k}/m_{k})$, which 
contradicts the assumption that 
$f(M_{1}/p_{t[I]}^{1},\ldots,M_{k}/p_{t[I]}^{k}) <f(p_{t[I]}^{1}/m_{1},  \ldots,p_{t[I]}^{k}/m_{k})$.}.~Assume~for simplicity that 
${\cal J}=\{1,\ldots,u\}$ for $u \geq 1$. 
By setting $p_{t[I]}^{j} = m_{j}$ for each $j \in {\cal J}$,~we~have~that 
\[
f\left(\frac{M_{1}}{m_{1}},\ldots,\frac{M_{u}}{m_{u}},
\frac{M_{u+1}}{p_{t[I]}^{u+1}},\ldots,
\frac{M_{k}}{p_{t[I]}^{k}}\right) \geq 
f\left(1,\ldots,1,\frac{p_{t[I]}^{u+1}}{m_{u+1}},
\ldots,\frac{p_{t[I]}^{k}}{m_{k}}\right).
\]
Since $f$ is monotone and continuous, 
there exist $q_{t[I]}^{1} \in  [m_{1},p_{t[I]}^{1}], \ldots,
q_{t[I]}^{u} \in  [m_{u},p_{t[I]}^{u}]$ such that 
\begin{eqnarray*}
f\left(\frac{M_{1}}{p_{t[I]}^{1}},\ldots,\frac{M_{k}}{p_{t[I]}^{k}}\right) & \leq & 
f\left(\frac{M_{1}}{q_{t[I]}^{1}},\ldots,\frac{M_{u}}{q_{t[I]}^{u}},
\frac{M_{u+1}}{p_{t[I]}^{u+1},} \ldots,\frac{M_{k}}{p_{t[I]}^{k}}\right)\\
& = & f\left(\frac{q_{t[I]}^{1}}{m_{1}},\ldots,\frac{q_{t[I]}^{u}}{m_{u}},
\frac{p_{t[I]}^{u+1}}{m_{u+1}},\ldots,\frac{p_{t[I]}^{k}}{m_{k}}\right)
\leq f\left(\frac{p_{t[I]}^{1}}{m_{1}},\ldots,\frac{p_{t[I]}^{k}}{m_{k}}\right).
\end{eqnarray*}
Then it turns out that $(q_{t[I]}^{1},\ldots,q_{t[I]}^{u},p_{t[I]}^{u+1},
\ldots,p_{t[I]}^{k}) \in {\cal S}_{f}^{k}$ and it follows that 
\[
f\left(\frac{M_{1}}{p_{t[I]}^{1}},\ldots,\frac{M_{k}}{p_{t[I]}^{k}}\right) 
\leq f\left(\frac{M_{1}}{q_{t[I]}^{1}},\ldots,\frac{M_{u}}{q_{t[I]}^{u}},
\frac{M_{u+1}}{p_{t[I]}^{u+1},} \ldots,\frac{M_{k}}{p_{t[I]}^{k}}\right )
\leq z_{f}^{k}. 
\]

For each $I = ( \vec{p}_{1},\ldots,\vec{p}_{T}) 
\in {\cal I}_{\rm rej}$, 
the algorithm $\msc{bpp}_{k}$ rejects a price vector 
$\vec{p}_{t}$~for~every~$t \in [1,T]$, and settles in the 
minimum price vector $\vec{p}_{\rm min} = ( m_{1},\ldots,m_{k})$, but 
at time $\tau[I] \in [1,T]$,~the optimal offline algorithm 
$\msc{opt}_{k}$ can accept a price vector 
$\vec{p}_{\tau[I]}=(p_{\tau[I]}^{1},\ldots,p_{\tau[I]}^{k})$ satisfying that 
$f(p_{\tau[I]}^{1}/m_{1},\ldots,p_{\tau[I]}^{k}/m_{k})=
\max_{t \in [1,T]} f(p_{t}^{1}/m_{1},\ldots,p_{t}^{k}/m_{k})$. So  
from~Equation~(\ref{eq-rej-bound}),~we~have~that 
\[
{\cal CR}^{f}(\msc{bpp}_{k};I) = 
f\left(
\frac{p_{\tau[I]}^{1}}{m_{1}}, \ldots,
\frac{p_{\tau[I]}^{k}}{m_{k}} \right). 
\]
We show that $f(p_{\tau[I]}^{1}/m_{1},\ldots,p_{\tau[I]}^{k}/m_{k})
\leq z_{f}^{k}$. Since the algorithm 
$\msc{bpp}_{k}$ rejects a price vector
$\vec{p}_{t}$ for every $t \in [1,T]$, 
it is immediate that 
$f(M_{1}/p_{\tau[I]}^{1},\ldots,M_{k}/p_{\tau[I]}^{k}) > 
f(p_{\tau[I]}^{1}/m_{1},\ldots,p_{\tau[I]}^{k}/m_{k})$ by definition. 
Let ${\cal H} = \{h \in [1,k]: 
M_{h}/p_{\tau[I]}^{h}\geq p_{\tau[I]}^{h}/m_{h}\}$. 
We claim 
that ${\cal H} \neq \emptyset$\footnote{~By contradiction. 
If ${\cal H}=\emptyset$, 
then $M_{i}/p_{\tau[I]}^{i} < p_{\tau[I]}^{i}/m_{i}$ for each 
$i \in [1,k]$. Since the function $f: {\bf R}^{k} \to {\bf R}$ is 
monotone, we have that 
$f(M_{1}/p_{\tau[I]}^{1},\ldots,M_{k}/p_{\tau[I]}^{k}) \leq  
f(p_{\tau[I]}^{1}/m_{1},\ldots,p_{\tau[I]}^{k}/m_{k})$, 
which~contradicts~the~assumption that $f(M_{1}/p_{\tau[I]}^{1},\ldots,
M_{k}/p_{\tau[I]}^{k})>f(p_{\tau	[I]}^{1}/m_{1},\ldots,p_{\tau[I]}^{k}/m_{k})$.}. 
For~simplicity, we  
assume that ${\cal H}=\{1,\ldots,v\}$ for $v \geq 1$. 
By setting $p_{\tau[I]}^{h} = M_{h}$ for each $h \in {\cal H}$,~we~have~that 
\[
f\left(1,\ldots,1,\frac{M_{v+1}}{p_{\tau[I]}^{v+1}},\ldots,
\frac{M_{k}}{p_{\tau[I]}^{k}}\right) \leq 
f\left(\frac{M_{1}}{m_{1}},\ldots,\frac{M_{v}}{m_{v}},
\frac{p_{\tau[I]}^{v+1}}{m_{v+1}},
\ldots,\frac{p_{\tau[I]}^{k}}{m_{k}}\right).
\]
Since $f$ is monotone and continuous, 
there exist $q_{\tau[I]}^{1} \in  [p_{\tau[I]}^{1},M_{1}], \ldots,
q_{\tau[I]}^{v} \in  [p_{\tau[I]}^{v},M_{v}]$ ~such~that 
\begin{eqnarray*}
f\left(\frac{p_{\tau[I]}^{1}}{m_{1}},\ldots,\frac{p_{\tau[I]}^{k}}{m_{k}}\right) 
& \leq & 
f\left(\frac{q_{\tau[I]}^{1}}{m_{1}},\ldots,\frac{q_{\tau[I]}^{v}}{m_{v}},
\frac{p_{\tau[I]}^{v+1}}{m_{v+1}}, \ldots,\frac{p_{\tau[I]}^{k}}{m_{k}}\right)\\
& = & f\left(\frac{M_{1}}{q_{\tau[I]}^{1}},\ldots,\frac{M_{v}}{q_{\tau[I]}^{v}},
\frac{M_{v+1}}{p_{\tau[I]}^{v+1}},\ldots,\frac{M_{k}}{p_{\tau[I]}^{k}}\right)
\leq f\left(\frac{M_{1}}{p_{\tau[I]}^{1}},\ldots,\frac{M_{k}}{p_{\tau[I]}^{k}}\right).
\end{eqnarray*}
Then it turns out that $(q_{\tau[I]}^{1},\ldots,q_{\tau[I]}^{v},p_{\tau[I]}^{v+1},
\ldots,p_{\tau[I]}^{k}) \in {\cal S}_{f}^{k}$ and it follows that 
\[
f\left(\frac{p_{\tau[I]}^{1}}{m_{1}},\ldots,\frac{p_{\tau[I]}^{k}}{m_{k}}\right) 
\leq 
f\left(\frac{q_{\tau[I]}^{1}}{m_{1}},\ldots,\frac{q_{\tau[I]}^{v}}{m_{v}},
\frac{p_{\tau[I]}^{v+1}}{m_{v+1}}, \ldots,\frac{p_{\tau[I]}^{k}}{m_{k}}\right)
\leq z_{f}^{k}. 
\]

Note that ${\cal I}_{\rm acc}
\cap {\cal I}_{\rm rej}=\emptyset$ and 
${\cal I}_{\rm acc} \cup 
{\cal I}_{\rm acc} = {\cal I}$. 
Thus for any $I \in {\cal I}$, we have that 
${\cal CR}^{f}(\msc{bpp}_{k};I) \leq z_{f}^{k}$ and we can conclude that 
${\cal CR}^{f}(\msc{bpp}_{k}) = \sup_{I \in {\cal I}} {\cal CR}^{f}(\msc{bpp}_{k};I) 
\leq z_{f}^{k}$. \BQED
\begin{theorem} \label{thm-unified-lower}
Let $\msc{alg}_{k}$ be an arbitrary online algorithm for 
the multi-objective $(k$-objective$)$ time series search problem. 
If all of $\msc{itv}_{1}=[m_{1},M_{1}],\ldots,\msc{itv}_{k}=[m_{k},M_{k}]$ 
are real intervals,~then  
for any monotone continuous function $f: {\bf R}^{k} \to {\bf R}$ and 
any integer $k \geq 1$, 
${\cal CR}^{f}(\msc{alg}_{k}) \geq z_{f}^{k}$. 
\end{theorem}
{\bf Proof:} Let $\msc{alg}_{k}$ be an arbitrarily online algorithm and 
$(x_{1}^{*},\ldots,x_{k}^{*}) \in {\cal S}_{f}^{k}$ be a price vector such that 
$z_{f}^{k} = f(M_{1}/x_{1}^{*},\ldots,M_{k}/x_{k}^{*})$. 
The adversary reveals a price vector 
$\vec{p}=(x_{1}^{*},\ldots,x_{k}^{*})$.~If~the~algorithm $\msc{alg}_{k}$ accepts 
$\vec{p}$, then the adversary reveals 
another price vector $\vec{p}_{\rm max}=(M_{1},\ldots,M_{k})$ and accepts $\vec{p}_{\rm max}$. Let $I=(\vec{p},\vec{p}_{\rm max})$ be an input sequence. 
Then we have that 
\[
{\cal CR}^{f}(\msc{alg}_{k};I) = 
f\left(\frac{M_{1}}{x_{1}^{*}},\ldots,
\frac{M_{k}}{x_{k}^{*}}\right) = z_{f}^{k}. 
\]
If the algorithm $\msc{alg}_{k}$ rejects $\vec{p}$, then 
the adversary accepts $\vec{p}$ but reveals no further 
price vectors until the algorithm $\msc{alg}_{k}$ settles in the minimum 
price vector $\vec{p}_{\rm min}=(m_{1},\ldots,m_{k})$. Let $J = (\vec{p})$ 
be an input sequence. Note that $
z_{f}^{k}=f(x_{1}^{*}/m_{1},\ldots,x_{k}^{*}/m_{k})$.  Then we also have that 
\[
{\cal CR}^{f}(\msc{alg}_{k};J) = 
f\left(\frac{x_{1}^{*}}{m_{1}},\ldots,\frac{x_{k}^{*}}{m_{k}}\right) 
= z_{f}^{k}.
\] 
Thus for any online algorithm $\msc{alg}_{k}$, it follows that 
${\cal CR}^{f}(\msc{alg}_{k})= \sup_{I \in {\cal I}} {\cal CR}^{f}(\msc{alg}_{k};I) 
\geq z_{f}^{k}$.~\BQED\medskip

From Theorems \ref{thm-unified-upper} and \ref{thm-unified-lower}, 
we immediately have the following result. 
\begin{corollary} \label{cor-unified}
If all of $\msc{itv}_{1}=[m_{1},M_{1}],\ldots,\msc{itv}_{k}=[m_{k},M_{k}]$ 
are real intervals, then 
for any monotone continuous function $f: {\bf R}^{k} \to {\bf R}$ and 
any integer $k\geq 1$, ${\cal CR}^{f}(\msc{bpp}_{k}) = z_{f}^{k}$. 
\end{corollary}
%
%====================================================
\subsection{Discussions} \label{subsec-discussion}
%====================================================
%
As mentioned in Subsection \ref{subsec-previous}, 
El-Yaniv, et al. \cite{Eetal} presented the 
algorithm {\sc rpp} (reservation price policy) 
for the single-objective time series search problem (see Figure \ref{fig-rpp}). 
We refer to $p^{*}$~as the {\it reservation price\/}, where 
$p^{*}$ is the solution of $M/p=p/m$. 
\begin{figure*}[h]
\begin{center}
\fbox{
\begin{minipage}{6.05cm}\smallskip
\begin{tabular}{l}
{\bf for} $t=1,2,\ldots, T$ {\bf do}\\
   $|$\\[-0.25cm]
   $|$  Accept $p_{t}$ if 
$p_{t}\geq p^{*}=\sqrt{Mm}$.\\[-0.25cm]
   $|$\\
{\bf end}
\end{tabular}
\end{minipage}
}
\caption{Reservation Price Policy: $\msc{rpp}$} \label{fig-rpp}
\end{center}
\end{figure*}

\noindent For the monotone continuous functions 
$f_{1}$, $f_{2}$, and $f_{3}$, we have that 
$f_{1}(x)=f_{2}(x)=f_{3}(x)=x$ if $k=1$, and 
the algorithm $\msc{bpp}_{1}$ coincides with the algorithm {\sc rpp} 
with respect to the functions 
$f_{1}$, $f_{2}$, and $f_{3}$, however, 
this is not necessarily 
the case for any nondecreasing\footnote{~For $k=1$, it is obvious that 
any monotone continuous 
function $f:{\bf R}\to {\bf R}$ is nondecreasing and continuous.} 
continuous~functions 
$f: {\bf R}\to {\bf R}$. Let us consider the following 
nondecreasing continuous function $g: {\bf R}\to {\bf R}$. 

\begin{figure*}[h]
\begin{center}
%
%WinTpicVersion4.23
\unitlength 0.1in
\begin{picture}( 44.1000, 34.0000)( -4.2000,-36.3500)
% VECTOR 1 0 3 0 Black White
% 2 1000 3300 1000 400
% 
{\color[named]{Black}{%
\special{pn 13}%
\special{pa 1000 3300}%
\special{pa 1000 400}%
\special{fp}%
\special{sh 1}%
\special{pa 1000 400}%
\special{pa 980 468}%
\special{pa 1000 454}%
\special{pa 1020 468}%
\special{pa 1000 400}%
\special{fp}%
}}%
% VECTOR 1 0 3 0 Black White
% 2 900 3200 3800 3200
% 
{\color[named]{Black}{%
\special{pn 13}%
\special{pa 900 3200}%
\special{pa 3800 3200}%
\special{fp}%
\special{sh 1}%
\special{pa 3800 3200}%
\special{pa 3734 3180}%
\special{pa 3748 3200}%
\special{pa 3734 3220}%
\special{pa 3800 3200}%
\special{fp}%
}}%
% LINE 2 0 3 0 Black White
% 2 1400 3250 1400 3150
% 
{\color[named]{Black}{%
\special{pn 8}%
\special{pa 1400 3250}%
\special{pa 1400 3150}%
\special{fp}%
}}%
% LINE 2 0 3 0 Black White
% 2 950 2800 1050 2800
% 
{\color[named]{Black}{%
\special{pn 8}%
\special{pa 950 2800}%
\special{pa 1050 2800}%
\special{fp}%
}}%
% STR 2 0 3 0 Black White
% 4 1400 3250 1400 3350 5 0 0 0
% $1$
\put(14.0000,-33.5000){\makebox(0,0){$1$}}%
% STR 2 0 3 0 Black White
% 4 900 2700 900 2800 5 0 0 0
% $1$
\put(9.0000,-28.0000){\makebox(0,0){$1$}}%
% LINE 2 0 3 0 Black White
% 2 2000 3250 2000 3150
% 
{\color[named]{Black}{%
\special{pn 8}%
\special{pa 2000 3250}%
\special{pa 2000 3150}%
\special{fp}%
}}%
% LINE 2 0 3 0 Black White
% 2 3600 3250 3600 3150
% 
{\color[named]{Black}{%
\special{pn 8}%
\special{pa 3600 3250}%
\special{pa 3600 3150}%
\special{fp}%
}}%
% LINE 2 0 3 0 Black White
% 2 2600 3250 2600 3150
% 
{\color[named]{Black}{%
\special{pn 8}%
\special{pa 2600 3250}%
\special{pa 2600 3150}%
\special{fp}%
}}%
% LINE 2 0 3 0 Black White
% 2 950 2200 1050 2200
% 
{\color[named]{Black}{%
\special{pn 8}%
\special{pa 950 2200}%
\special{pa 1050 2200}%
\special{fp}%
}}%
% LINE 2 0 3 0 Black White
% 2 950 600 1050 600
% 
{\color[named]{Black}{%
\special{pn 8}%
\special{pa 950 600}%
\special{pa 1050 600}%
\special{fp}%
}}%
% LINE 2 2 3 0 Black White
% 2 1400 2800 1000 2800
% 
{\color[named]{Black}{%
\special{pn 8}%
\special{pa 1400 2800}%
\special{pa 1000 2800}%
\special{dt 0.045}%
}}%
% LINE 2 2 3 0 Black White
% 2 2000 3200 2000 2200
% 
{\color[named]{Black}{%
\special{pn 8}%
\special{pa 2000 3200}%
\special{pa 2000 2200}%
\special{dt 0.045}%
}}%
% LINE 2 2 3 0 Black White
% 2 2000 2200 1000 2200
% 
{\color[named]{Black}{%
\special{pn 8}%
\special{pa 2000 2200}%
\special{pa 1000 2200}%
\special{dt 0.045}%
}}%
% LINE 2 2 3 0 Black White
% 2 2600 3200 2600 1600
% 
{\color[named]{Black}{%
\special{pn 8}%
\special{pa 2600 3200}%
\special{pa 2600 1600}%
\special{dt 0.045}%
}}%
% LINE 2 2 3 0 Black White
% 2 3600 3200 3600 600
% 
{\color[named]{Black}{%
\special{pn 8}%
\special{pa 3600 3200}%
\special{pa 3600 600}%
\special{dt 0.045}%
}}%
% LINE 2 2 3 0 Black White
% 2 3600 600 1000 600
% 
{\color[named]{Black}{%
\special{pn 8}%
\special{pa 3600 600}%
\special{pa 1000 600}%
\special{dt 0.045}%
}}%
% LINE 2 2 3 0 Black White
% 2 1000 3200 3750 450
% 
{\color[named]{Black}{%
\special{pn 8}%
\special{pa 1000 3200}%
\special{pa 3750 450}%
\special{dt 0.045}%
}}%
% LINE 1 0 3 0 Black White
% 2 2600 1600 3600 600
% 
{\color[named]{Black}{%
\special{pn 13}%
\special{pa 2600 1600}%
\special{pa 3600 600}%
\special{fp}%
}}%
% STR 2 0 3 0 Black White
% 4 750 2100 750 2200 5 0 0 0
% $\sqrt{\frac{M}{m}}$
\put(7.5000,-22.0000){\makebox(0,0){$\sqrt{\frac{M}{m}}$}}%
% STR 2 0 3 0 Black White
% 4 2600 3350 2600 3450 5 0 0 0
% $c\sqrt{\frac{M}{m}}$
\put(26.0000,-34.5000){\makebox(0,0){$c\sqrt{\frac{M}{m}}$}}%
% STR 2 0 3 0 Black White
% 4 3600 3350 3600 3450 5 0 0 0
% $\frac{M}{m}$
\put(36.0000,-34.5000){\makebox(0,0){$\frac{M}{m}$}}%
% STR 2 0 3 0 Black White
% 4 750 1500 750 1600 5 0 0 0
% $c \sqrt{\frac{M}{m}}$
\put(7.5000,-16.0000){\makebox(0,0){$c \sqrt{\frac{M}{m}}$}}%
% STR 2 0 3 0 Black White
% 4 800 500 800 600 5 0 0 0
% $\frac{M}{m}$
\put(8.0000,-6.0000){\makebox(0,0){$\frac{M}{m}$}}%
% STR 2 0 3 0 Black White
% 4 1000 200 1000 300 5 0 0 0
% $g(x)$
\put(10.0000,-3.0000){\makebox(0,0){$g(x)$}}%
% STR 2 0 3 0 Black White
% 4 3900 3100 3900 3200 5 0 0 0
% $x$
\put(39.0000,-32.0000){\makebox(0,0){$x$}}%
% VECTOR 2 0 3 0 Black White
% 2 2400 1000 2400 1000
% 
{\color[named]{Black}{%
\special{pn 8}%
\special{pa 2400 1000}%
\special{pa 2400 1000}%
\special{fp}%
}}%
% VECTOR 2 0 3 0 Black White
% 2 2400 1000 2800 1400
% 
{\color[named]{Black}{%
\special{pn 8}%
\special{pa 2400 1000}%
\special{pa 2800 1400}%
\special{fp}%
\special{sh 1}%
\special{pa 2800 1400}%
\special{pa 2768 1340}%
\special{pa 2762 1362}%
\special{pa 2740 1368}%
\special{pa 2800 1400}%
\special{fp}%
}}%
% STR 2 0 3 0 Black White
% 4 2350 800 2350 900 5 0 0 0
% $g(x)$
\put(23.5000,-9.0000){\makebox(0,0){$g(x)$}}%
% LINE 2 0 3 0 Black White
% 2 960 1600 1060 1600
% 
{\color[named]{Black}{%
\special{pn 8}%
\special{pa 960 1600}%
\special{pa 1060 1600}%
\special{fp}%
}}%
% LINE 1 0 3 0 Black White
% 2 1400 1600 2600 1600
% 
{\color[named]{Black}{%
\special{pn 13}%
\special{pa 1400 1600}%
\special{pa 2600 1600}%
\special{fp}%
}}%
% LINE 2 2 3 0 Black White
% 2 1400 1600 1400 3200
% 
{\color[named]{Black}{%
\special{pn 8}%
\special{pa 1400 1600}%
\special{pa 1400 3200}%
\special{dt 0.045}%
}}%
% LINE 2 2 3 0 Black White
% 2 1000 1600 1400 1600
% 
{\color[named]{Black}{%
\special{pn 8}%
\special{pa 1000 1600}%
\special{pa 1400 1600}%
\special{dt 0.045}%
}}%
% STR 2 0 3 0 Black White
% 4 1600 3600 1600 3700 5 0 0 0
% $\frac{1}{c}\sqrt{\frac{M}{m}}$
\put(16.0000,-37.0000){\makebox(0,0){$\frac{1}{c}\sqrt{\frac{M}{m}}$}}%
% STR 2 0 3 0 Black White
% 4 400 2500 400 2600 5 0 0 0
% $\frac{1}{c}\sqrt{\frac{M}{m}}$
\put(4.0000,-26.0000){\makebox(0,0){$\frac{1}{c}\sqrt{\frac{M}{m}}$}}%
% LINE 2 0 3 0 Black White
% 2 950 2600 1050 2600
% 
{\color[named]{Black}{%
\special{pn 8}%
\special{pa 950 2600}%
\special{pa 1050 2600}%
\special{fp}%
}}%
% LINE 2 0 3 0 Black White
% 2 1600 3250 1600 3150
% 
{\color[named]{Black}{%
\special{pn 8}%
\special{pa 1600 3250}%
\special{pa 1600 3150}%
\special{fp}%
}}%
% LINE 2 2 3 0 Black White
% 2 1600 3200 1600 2600
% 
{\color[named]{Black}{%
\special{pn 8}%
\special{pa 1600 3200}%
\special{pa 1600 2600}%
\special{dt 0.045}%
}}%
% LINE 2 2 3 0 Black White
% 2 1600 2600 1000 2600
% 
{\color[named]{Black}{%
\special{pn 8}%
\special{pa 1600 2600}%
\special{pa 1000 2600}%
\special{dt 0.045}%
}}%
% VECTOR 2 0 3 0 Black White
% 2 700 2600 870 2600
% 
{\color[named]{Black}{%
\special{pn 8}%
\special{pa 700 2600}%
\special{pa 870 2600}%
\special{fp}%
\special{sh 1}%
\special{pa 870 2600}%
\special{pa 804 2580}%
\special{pa 818 2600}%
\special{pa 804 2620}%
\special{pa 870 2600}%
\special{fp}%
}}%
% VECTOR 2 0 3 0 Black White
% 2 1600 3500 1600 3300
% 
{\color[named]{Black}{%
\special{pn 8}%
\special{pa 1600 3500}%
\special{pa 1600 3300}%
\special{fp}%
\special{sh 1}%
\special{pa 1600 3300}%
\special{pa 1580 3368}%
\special{pa 1600 3354}%
\special{pa 1620 3368}%
\special{pa 1600 3300}%
\special{fp}%
}}%
% STR 2 0 3 0 Black White
% 4 2000 3350 2000 3450 5 0 0 0
% $\sqrt{\frac{M}{m}}$
\put(20.0000,-34.5000){\makebox(0,0){$\sqrt{\frac{M}{m}}$}}%
\end{picture}%

\end{center}
\caption{Counterexample for Nondecreasing Continuous Function 
$g: {\bf R}\to {\bf R}$} \label{fig-counter}
\end{figure*}
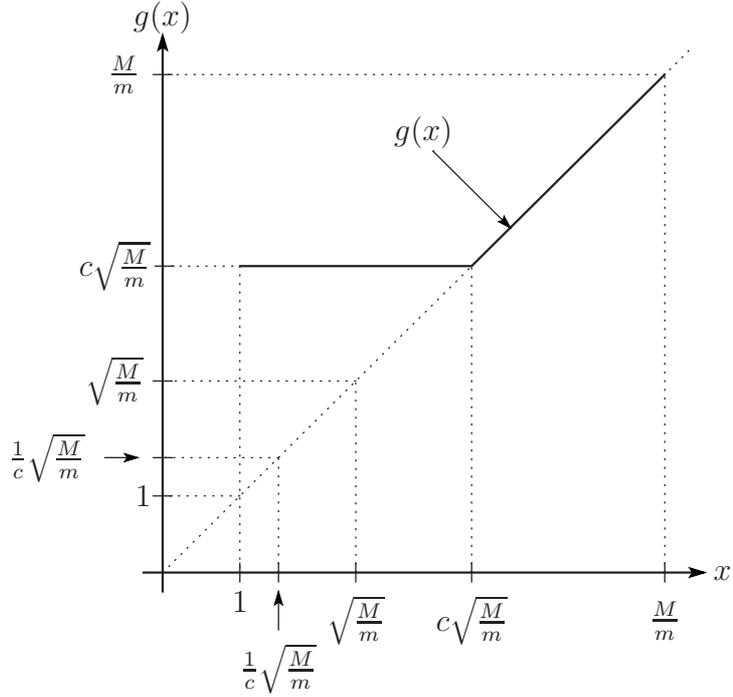

\noindent 
From the assumption that $0 < m < M $, it follows 
that $M/m>1$ and we can take 
any constant~$c$ such that $1 < c < \sqrt{M/m}$. 
Then it is immediate that 
\begin{eqnarray*}
g(M/p) > g(p/m) & & \mbox{for $m \leq p < \sqrt{Mm}/c$};\\
g(M/p) = g(p/m) & & \mbox{for $\sqrt{Mm}/c\leq p \leq c \sqrt{Mm}$};\\
g(M/p) <  g(p/m) & & \mbox{for $c \sqrt{Mm} < p \leq M$}.
\end{eqnarray*}
Thus the algorithm $\msc{bpp}_{1}$ does not coincide with the algorithm 
{\sc rpp} \cite{Eetal} with respect to the nondecreasing continuous 
(equivalently monotone) 
function $g: {\bf R}\to {\bf R}$ in Figure \ref{fig-counter}. 
%
%========================================================
\section{Analysis for Competitive Ratio} \label{sec-ratio}
%========================================================
%
For the case that all of 
$\msc{itv}_{1}=[m_{1},M_{1}],\ldots,
\msc{itv}_{k}=[m_{k},M_{k}]$ are real intervals, 
Corollary~\ref{cor-unified} gives the best possible value 
of the competitive ratio for the multi-objective time~series search problem 
with respect to any monotone {\it continuous\/} 
function $f$. 
In this section, we assume that 
all of $\msc{itv}_{1}=[m_{1},M_{1}],\ldots,
\msc{itv}_{k}=[m_{k},M_{k}]$ are real intervals,~and~derive~the~best 
possible values of the competitive ratio for the multi-objective time
series search problem with respect to the monotone
functions $f_{1}$, $f_{2}$, $f_{3}$, and 
$f_{4}$ in Subsections \ref{subsec-worst}, 
\ref{subsec-arithmetic}, 
\ref{subsec-geometric}, and 
\ref{subsec-best}, respectively. 
%
%=================================================
\subsection{Worst Component Competitive Ratio} 
\label{subsec-worst}
%=================================================
%
In this subsection, we show that ${\cal CR}^{f_{1}}(\msc{bpp}_{k})=z_{f_{1}}^{k}=
\max\{\sqrt{M_{1}/m_{1}},M_{2}/m_{2}\}$. 
This implies that the algorithm {\sc rpp-high} \cite[Algorithm 1]{TIS} can 
be regarded as a special case of the 
algorithm $\msc{bpp}_{k}$ with respect to the function 
$f_{1}(c_{1},\ldots,c_{k})=\max(c_{1},\ldots,c_{k})$. 
For the function $f_{1}$, let 
\begin{eqnarray*}
{\cal S}_{f_{1}}^{k} & = & \left\{(x_{1},\ldots,x_{k})\in 
I_{1} \times \cdots \times I_{k}: 
\max \left(\frac{M_{1}}{x_{1}}, \ldots, 
\frac{M_{k}}{x_{k}}\right)=\max \left(\frac{x_{1}}{m_{1}}, \ldots, 
\frac{x_{k}}{m_{k}}\right)\right\};\\
z_{f_{1}}^{k} & = & \sup_{(x_{1},\ldots,x_{k}) \in  {\cal S}_{f_{1}}^{k}}
\left[\max\left(\frac{M_{1}}{x_{1}},\ldots,\frac{M_{k}}{x_{k}}\right) \right].
\end{eqnarray*}
\begin{theorem} \label{thm-worst}
$z_{f_{1}}^{k}
=\max\{\sqrt{M_{1}/m_{1}},M_{2}/m_{2}\}$ 
for any integer $k \geq 2$. 
\end{theorem}
{\bf Proof:} Consider the following two cases: 
(1) $\sqrt{M_{1}/m_{1}} \geq M_{2}/m_{2}$ and 
(2) $\sqrt{M_{1}/m_{1}} < M_{2}/m_{2}$. 

For the case (1), 
we further consider the following three subcases: 
(1.1) $x_{1}> \sqrt{m_{1}M_{1}}$, (1.2) $x_{1}< \sqrt{m_{1}M_{1}}$, and 
(1.3) $x_{1}= \sqrt{m_{1}M_{1}}$.  For the subcase (1.1), we have that 
\begin{eqnarray*}
\left. 
\begin{array}{rcl}
\frac{M_{1}}{x_{1}} & < & \frac{M_{1}}{\sqrt{m_{1}M_{1}}} 
= \sqrt{\frac{M_{1}}{m_{1}}}\\
\frac{M_{2}}{x_{2}} & \leq & \frac{M_{2}}{m_{2}} \leq \sqrt{\frac{M_{1}}{m_{1}}}\\
  & \vdots & \\
\frac{M_{k}}{x_{k}} & \leq & \frac{M_{k}}{m_{k}} \leq \frac{M_{2}}{m_{2}}\leq 
\sqrt{\frac{M_{1}}{m_{1}}}
\end{array} \right\} & \Rightarrow & 
f_{1}\left(\frac{M_{1}}{x_{1}},\ldots,\frac{M_{k}}{x_{k}}\right) =
\max\left( \frac{M_{1}}{x_{1}},\ldots,\frac{M_{k}}{x_{k}}\right)
\leq  \sqrt{\frac{M_{1}}{m_{1}}};\\
\left. 
\begin{array}{rcl}
\frac{x_{1}}{m_{1}} & > & \frac{\sqrt{m_{1}M_{1}}}{m_{1}} 
= \sqrt{\frac{M_{1}}{m_{1}}}\\
\frac{x_{2}}{m_{2}} & \leq & \frac{M_{2}}{m_{2}}\leq \sqrt{\frac{M_{1}}{m_{1}}}\\
  & \vdots & \\
\frac{x_{k}}{m_{k}} & \leq & \frac{M_{k}}{m_{k}} \leq \frac{M_{2}}{m_{2}}\leq 
\sqrt{\frac{M_{1}}{m_{1}}}
\end{array} \right\} & \Rightarrow & 
f_{1}\left(\frac{x_{1}}{m_{1}},\ldots,\frac{x_{k}}{m_{k}}\right) = 
\max\left(\frac{x_{1}}{m_{1}},\ldots,\frac{x_{k}}{m_{k}}\right) >    
\sqrt{\frac{M_{1}}{m_{1}}}. 
\end{eqnarray*}
Thus $f_{1}(M_{1}/x_{1},\ldots,M_{k}/x_{k}) < 
f_{1}(x_{1}/m_{1},\ldots,x_{k}/m_{k})$. 
For the subcase (1.2), we have that 
\begin{eqnarray*}
\left. 
\begin{array}{rcl}
\frac{M_{1}}{x_{1}} & > & \frac{M_{1}}{\sqrt{m_{1}M_{1}}} 
= \sqrt{\frac{M_{1}}{m_{1}}}\\
\frac{M_{2}}{x_{2}} & \leq & \frac{M_{2}}{m_{2}} \leq \sqrt{\frac{M_{1}}{m_{1}}}\\
  & \vdots & \\
\frac{M_{k}}{x_{k}} & \leq & \frac{M_{k}}{m_{k}} \leq \frac{M_{2}}{m_{2}}
\leq \sqrt{\frac{M_{1}}{m_{1}}}
\end{array} \right\} & \Rightarrow & 
f_{1}\left(\frac{M_{1}}{x_{1}},\ldots,\frac{M_{k}}{x_{k}}\right) = 
\max \left(\frac{M_{1}}{x_{1}},\ldots,\frac{M_{k}}{x_{k}}\right) >  
\sqrt{\frac{M_{1}}{m_{1}}};\\
\left. 
\begin{array}{rcl}
\frac{x_{1}}{m_{1}} & < & \frac{\sqrt{m_{1}M_{1}}}{m_{1}} 
= \sqrt{\frac{M_{1}}{m_{1}}}\\
\frac{x_{2}}{m_{2}} & \leq & \frac{M_{2}}{m_{2}}\leq \sqrt{\frac{M_{1}}{m_{1}}}\\
  & \vdots & \\
\frac{x_{k}}{m_{k}} & \leq & \frac{M_{k}}{m_{k}} \leq \frac{M_{2}}{m_{2}}\leq 
\sqrt{\frac{M_{1}}{m_{1}}}
\end{array} \right\} & \Rightarrow & 
f_{1}\left(\frac{x_{1}}{m_{1}},\ldots,\frac{x_{k}}{m_{k}}\right) = 
\max \left(\frac{x_{1}}{m_{1}},\ldots,\frac{x_{k}}{m_{k}}\right) \leq  
\sqrt{\frac{M_{1}}{m_{1}}}. 
\end{eqnarray*}
Thus $f_{1}(M_{1}/x_{1},\ldots,M_{k}/x_{k}) > 
f_{1}(x_{1}/m_{1},\ldots,x_{k}/m_{k})$. 
For the subcase (1.3), we have that 
\begin{eqnarray*}
\left. 
\begin{array}{rcl}
\frac{M_{1}}{x_{1}} & = & \frac{M_{1}}{\sqrt{m_{1}M_{1}}} 
= \sqrt{\frac{M_{1}}{m_{1}}}\\
\frac{M_{2}}{x_{2}} & \leq & \frac{M_{2}}{m_{2}} \leq \sqrt{\frac{M_{1}}{m_{1}}}\\
  & \vdots & \\
\frac{M_{k}}{x_{k}} & \leq & \frac{M_{k}}{m_{k}} \leq \frac{M_{2}}{m_{2}}
\leq \sqrt{\frac{M_{1}}{m_{1}}}
\end{array} \right\} & \Rightarrow & 
f_{1}\left(\frac{M_{1}}{x_{1}},\ldots,\frac{M_{k}}{x_{k}}\right) =
\max \left(\frac{M_{1}}{x_{1}},\ldots,\frac{M_{k}}{x_{k}}\right) 
=\sqrt{\frac{M_{1}}{m_{1}}}; \\
\left. 
\begin{array}{rcl}
\frac{x_{1}}{m_{1}} & = & \frac{\sqrt{m_{1}M_{1}}}{m_{1}} 
= \sqrt{\frac{M_{1}}{m_{1}}}\\
\frac{x_{2}}{m_{2}} & \leq & \frac{M_{2}}{m_{2}}\leq \sqrt{\frac{M_{1}}{m_{1}}}\\
  & \vdots & \\
\frac{x_{k}}{m_{k}} & \leq & \frac{M_{k}}{m_{k}} \leq \frac{M_{2}}{m_{2}}\leq 
\sqrt{\frac{M_{1}}{m_{1}}}
\end{array} \right\} & \Rightarrow & 
f_{1}\left(\frac{x_{1}}{m_{1}},\ldots,\frac{x_{k}}{m_{k}}\right) =
\max \left(\frac{x_{1}}{m_{1}},\ldots,\frac{x_{k}}{m_{k}}\right) 
=\sqrt{\frac{M_{1}}{m_{1}}}. 
\end{eqnarray*}
Then for the case (1), 
we have that $z_{f_{1}}^{k} = \sqrt{M_{1}/m_{1}}$, which is achieved 
at any $\vec{p}=(x_{1},\ldots,x_{k}) \in 
[m_{1},M_{1}]\times \cdots \times [m_{k},M_{k}]$ such that 
$x_{1}=\sqrt{m_{1}M_{1}} \in [m_{1},M_{1}]$. 

For the case (2), we consider the 
following two subcases: 
(2.1) $x_{1} < M_{1}m_{2}/M_{2}$ and (2.2)~$x_{1} \geq M_{1}m_{2}/M_{2}$. 
Note that $m_{1} \leq M_{1}m_{2}/M_{2} \leq M_{1}$. 
For the subcase (2.1), we have that 
\begin{eqnarray*}
\left. 
\begin{array}{rcl}
\frac{M_{1}}{x_{1}} & > & \frac{M_{1}}{M_{1}}\frac{M_{2}}{m_{2}}
= \frac{M_{2}}{m_{2}}\\
\frac{M_{2}}{x_{2}} & \leq & \frac{M_{2}}{m_{2}}\\
  & \vdots & \\
\frac{M_{k}}{x_{k}} & \leq & \frac{M_{k}}{m_{k}} \leq \frac{M_{2}}{m_{2}}
\end{array} \right\} & \Rightarrow & 
f_{1}\left(\frac{M_{1}}{x_{1}},\ldots,\frac{M_{k}}{x_{k}}\right) = 
\max \left(\frac{M_{1}}{x_{1}},\ldots,\frac{M_{k}}{x_{k}}\right) 
> \frac{M_{2}}{m_{2}};\\
\left. 
\begin{array}{rcl}
\frac{x_{1}}{m_{1}} & < & \frac{M_{1}}{m_{1}}\frac{m_{2}}{M_{2}} < 
\left(\frac{M_{2}}{m_{2}}\right)^{2}\frac{m_{2}}{M_{2}} = \frac{M_{2}}{m_{2}}\\
\frac{x_{2}}{m_{2}} & \leq & \frac{M_{2}}{m_{2}}\\
  & \vdots & \\
\frac{x_{k}}{m_{k}} & \leq & \frac{M_{k}}{m_{k}} \leq \frac{M_{2}}{m_{2}}
\end{array} \right\} & \Rightarrow & 
f_{1}\left(\frac{x_{1}}{m_{1}},\ldots,\frac{x_{k}}{m_{k}}\right) = 
\max \left(\frac{x_{1}}{m_{1}},\ldots,\frac{x_{k}}{m_{k}}\right) 
\leq \frac{M_{2}}{m_{2}}. 
\end{eqnarray*}
Thus 
$f_{1}(M_{1}/x_{1},\ldots,M_{k}/x_{k}) > 
f_{1}(x_{1}/m_{1},\ldots,x_{k}/m_{k})$. For the subcase (2.2), 
we have that 
\[
\left.
\begin{array}{rcl}
\frac{M_{1}}{x_{1}} & \leq & \frac{M_{1}}{M_{1}}\frac{M_{2}}{m_{2}}
= \frac{M_{2}}{m_{2}}\\
\frac{M_{2}}{x_{2}} & \leq & \frac{M_{2}}{m_{2}}\\
  & \vdots & \\
\frac{M_{k}}{x_{k}} & \leq & \frac{M_{k}}{m_{k}} \leq \frac{M_{2}}{m_{2}}
\end{array} \right\} \Rightarrow
f_{1}\left(\frac{M_{1}}{x_{1}},\ldots,\frac{M_{k}}{x_{k}}\right) 
= \max\left(\frac{M_{1}}{x_{1}},\ldots,\frac{M_{k}}{x_{k}}\right) 
\leq \frac{M_{2}}{m_{2}},
\]
which implies that 
$z_{f_{1}}^{k}=\sup_{(x_{1},\ldots,x_{k})\in {\cal S}_{f_{1}}^{k}}
f_{1}(M_{1}/x_{1},\ldots,M_{k}/x_{k})
\leq M_{2}/m_{2}$. For the subcase~(2.2), 
we show that $z_{f_{2}}^{k}= M_{2}/m_{2}$. Let $x_{1}'=M_{1}m_{2}/M_{2}$. 
Since $M_{1}/m_{1}\geq M_{2}/m_{2}$, 
we have~that~$x_{1}' \in [m_{1},M_{1}]$, and 
from the assumption that $\sqrt{M_{1}/m_{1}}<M_{2}/m_{2}$, 
we have that $x_{1}'/m_{1}<M_{2}/m_{2}$.~So from the fact that 
$M_{2}/m_{2} \geq M_{i}/x_{i} \geq 1$ and 
$M_{2}/m_{2} \geq x_{i}/m_{i}$ for each $i \in [3,k]$, 
it follows that for $x_{1}'=M_{1}m_{2}/M_{2}$, $x_{2}'=M_{2} \in [m_{2},M_{2}]$, and 
any $x_{3} \in [m_{3},M_{3}],\ldots,x_{k}\in [m_{k}.,M_{k}]$, 
\begin{eqnarray*}
f_{1}\left(\frac{M_{1}}{x_{1}'},\frac{M_{2}}{x_{2}'},\frac{M_{3}}{x_{3}},
\ldots,\frac{M_{k}}{x_{k}}\right) 
& = & \max \left(\frac{M_{1}}{x_{1}'},\frac{M_{2}}{x_{2}'},\frac{M_{3}}{x_{3}},
\ldots,\frac{M_{k}}{x_{k}}\right)\\
& = & \max \left(\frac{M_{1}}{M_{1}}\cdot \frac{M_{2}}{m_{2}},
\frac{M_{2}}{M_{2}},\frac{M_{3}}{x_{3}},\ldots,\frac{M_{k}}{x_{k}}\right)\\
& = & \max \left(\frac{M_{2}}{m_{2}},
1,\frac{M_{3}}{x_{3}},\ldots,\frac{M_{k}}{x_{k}}\right) 
= \frac{M_{2}}{m_{2}};\\
f_{1}\left(\frac{x_{1}'}{m_{1}},\frac{x_{2}'}{m_{2}},\frac{x_{3}}{m_{3}},
\ldots,\frac{x_{k}}{m_{k}}\right) 
& = & \max \left(\frac{x_{1}'}{m_{1}},\frac{x_{2}'}{m_{2}},\frac{x_{3}}{m_{3}},
\ldots,\frac{x_{k}}{m_{k}}\right)\\
& = & \max \left(\frac{x_{1}'}{m_{1}},\frac{M_{2}}{m_{2}},
\frac{x_{3}}{m_{3}},\ldots,\frac{x_{k}}{m_{k}}\right)
= \frac{M_{2}}{m_{2}}. 
\end{eqnarray*}
Let $x_{1}'' = m_{1}M_{2}/m_{2}$. 
Since $M_{1}/m_{1}\geq M_{2}/m_{2}$, we have that 
$x_{1}''\in [m_{1},M_{1}]$, and 
from~the~assumption that $\sqrt{M_{1}/m_{1}} < M_{2}/m_{2}$, 
we also have that 
$x_{1}''\geq M_{1}m_{2}/M_{2}$ and $M_{1}/x_{1}''<M_{2}/m_{2}$.~So~from 
the fact that  
$M_{2}/m_{2} \geq M_{i}/x_{i}$ and 
$M_{2}/m_{2} \geq x_{i}/m_{i}\geq 1$ for each $i \in [3,k]$, 
it~follows~that~for~$x_{1}''=m_{1}M_{2}/m_{2}$, 
$x_{2}''=m_{2} \in [m_{2},M_{2}]$, and 
any $x_{3} \in [m_{3},M_{3}],\ldots,x_{k}\in [m_{k}.,M_{k}]$, 
\begin{eqnarray*}
f_{1}\left(\frac{M_{1}}{x_{1}''},\frac{M_{2}}{x_{2}''},\frac{M_{3}}{x_{3}},
\ldots,\frac{M_{k}}{x_{k}}\right) 
& = & \max \left(\frac{M_{1}}{x_{1}''}, 
\frac{M_{2}}{x_{2}''},\frac{M_{3}}{x_{3}},
\ldots,\frac{M_{k}}{x_{k}}\right)\\
& = & \max \left(\frac{M_{1}}{x_{1}''}, 
\frac{M_{2}}{m_{2}},\frac{M_{3}}{x_{3}},
\ldots,\frac{M_{k}}{x_{k}}\right) 
= \frac{M_{2}}{m_{2}};\\
f_{1}\left(\frac{x_{1}''}{m_{1}},\frac{x_{2}''}{m_{2}},\frac{x_{3}}{m_{3}},
\ldots,\frac{x_{k}}{m_{k}}\right) 
& = & \max \left(\frac{x_{1}''}{m_{1}},\frac{x_{2}''}{m_{2}},\frac{x_{3}}{m_{3}},
\ldots,\frac{x_{k}}{m_{k}}\right)\\
& = & \max \left(\frac{M_{2}}{m_{2}}\cdot \frac{m_{1}}{m_{1}},\frac{m_{2}}{m_{2}},
\frac{x_{3}}{m_{3}},\ldots,\frac{x_{k}}{m_{k}}\right)\\
& = & \max \left(\frac{M_{2}}{m_{2}},1, 
\frac{x_{3}}{m_{3}},\ldots,\frac{x_{k}}{m_{k}}\right) 
= \frac{M_{2}}{m_{2}}.
\end{eqnarray*}
Then for the case (2), 
we have that $z_{f_{1}}^{k} = M_{2}/m_{2}$, which is achieved 
at any $\vec{p}=(x_{1},\ldots,x_{k}) \in 
[m_{1},M_{1}]\times \cdots \times [m_{k},M_{k}]$ 
such that 
$x_{1}=M_{1}m_{2}/M_{2} \in [m_{1},M_{1}]$ and 
$x_{2} = M_{2} \in [m_{2},M_{2}]$~or~$x_{1}=m_{1}M_{2}/m_{2} \in 
[m_{1},M_{1}]$ and 
$x_{2} =m_{2} \in [m_{2},M_{2}]$. 

Since we have that $z_{f_{1}}^{k}=\sqrt{M_{1}/m_{1}}$ for the case (1) 
$\sqrt{M_{1}/m_{1}}\geq M_{2}/m_{2}$ and 
$z_{f_{1}}^{k}=M_{2}/m_{2}$~for the case (2)  
$\sqrt{M_{1}/m_{1}}< M_{2}/m_{2}$, we can conclude that 
$z_{f_{1}}^{k}=\max\{\sqrt{M_{1}/m_{1}},M_{2}/m_{2}\}$. 
\BQED\medskip

With respect to the function $f_{1}$, 
Tiedemann, et al. \cite{TIS} presented the algorithm 
{\sc rpp-high} and showed that 
${\cal CR}^{f_{1}}(\msc{rpp-high})=\max\{\sqrt{M_{1}/m_{1}}, 
M_{2}/m_{2}\}$ \cite[Theorems 1 and 2]{TIS}. 
By combining 
Corollary \ref{cor-unified} and Theorem \ref{thm-worst}, 
we have that 
${\cal CR}^{f_{1}}(\msc{bpp}_{k}) = z_{f_{1}}^{k}
=\max\{\sqrt{M_{1}/m_{2}},M_{2}/m_{2}\}$,~and 
this is another 
proof for the optimality on the worst component competitive ratio. 
%
%=========================================================
\subsection{Arithmetic Mean Component Competitive Ratio} 
\label{subsec-arithmetic}
%=========================================================
%
For $c_{1},\ldots,c_{k} \in {\bf R}$, 
let $f_{2}(c_{1},\ldots,c_{k})=(c_{1}+\cdots+c_{k})/k$. 
For the function 
$f_{2}: {\bf R}^{k} \to {\bf R}$, let 
\begin{eqnarray*}
{\cal S}_{f_{2}}^{k} & = & \left\{(x_{1},\ldots,x_{k})\in 
I_{1} \times \cdots \times I_{k}: 
\frac{1}{k}\left(\frac{M_{1}}{x_{1}}+ \cdots+
\frac{M_{k}}{x_{k}}\right)=\frac{1}{k}\left(\frac{x_{1}}{m_{1}}+ \cdots+
\frac{x_{k}}{m_{k}}\right)\right\};\\
z_{f_{2}}^{k} & = & \sup_{(x_{1},\ldots,x_{k}) \in {\cal S}_{f_{2}}^{k}} 
\frac{1}{k} \left(\frac{M_{1}}{x_{1}}+\cdots+\frac{M_{k}}{x_{k}}\right)
= \frac{1}{k} \sup_{(x_{1},\ldots,x_{k}) \in {\cal S}_{f_{2}}^{k}} 
\left(\frac{M_{1}}{x_{1}}+\cdots+\frac{M_{k}}{x_{k}}\right).
\end{eqnarray*}
With respect to the function $f_{2}$, it follows 
from Corollary \ref{cor-unified}  that ${\cal R}_{s}^{f_{2}}(\msc{bpp}_{k}) 
= z_{f_{2}}^{k}$. In general,~it would be difficult to explicitly 
represent $z_{f_{2}}^{k}$ 
by $m_{1},\ldots,m_{k}$ and $M_{1},\ldots,M_{k}$. 
So~we~consider~the case that $k=2$ and 
we give an explicit form of $z_{f_{2}}^{2}$ by 
$m_{1},m_{2}$ and $M_{1},M_{2}$. 
\begin{theorem} \label{thm-arithmetic}
With respect to the function $f_{2}$ for $k=2$, the following holds$:$
\[
z_{f_{2}}^{2}=\frac{1}{2}\left[\sqrt{
\left\{\frac{1}{2}
\left(\frac{M_{2}}{m_{2}}-1\right)\right\}^{2} 
+ \frac{M_{1}}{m_{1}}}+ \frac{1}{2}\left(\frac{M_{2}}{m_{2}}+1\right)\right].
\]
\end{theorem}
{\bf Proof:} Let $k=2$. Then 
${\cal S}_{f_{2}}^{2}$ and $z_{f_{2}}^{2}$ are given by 
\begin{eqnarray*}
{\cal S}_{f_{2}}^{2} & = & \left\{(x_{1},x_{2})\in I_{1} \times I_{2}: 
\frac{1}{2}\left(\frac{M_{1}}{x_{1}}+ \frac{M_{2}}{x_{2}}\right)
=\frac{1}{2}\left(\frac{x_{1}}{m_{1}}+ \frac{x_{2}}{m_{2}}\right)\right\};\\
& = & \left\{(x_{1},x_{2})\in I_{1} \times I_{2}: 
\frac{M_{1}}{x_{1}}- \frac{x_{1}}{m_{1}}=
- \left(\frac{M_{2}}{x_{2}}-\frac{x_{2}}{m_{2}}\right)\right\};\\
z_{f_{2}}^{2} & = & \sup_{(x_{1},x_{2}) \in {\cal S}_{f_{2}}^{2}}
\frac{1}{2} \left(\frac{M_{1}}{x_{1}}+\frac{M_{2}}{x_{2}}\right)
= \frac{1}{2} \sup_{(x_{1},x_{2}) \in {\cal S}_{f_{2}}^{2}}
\left(\frac{M_{1}}{x_{1}}+\frac{M_{2}}{x_{2}}\right)\\
& = & \frac{1}{2} \sup_{(x_{1},x_{2}) \in {\cal S}_{f_{2}}^{2}} 
\left\{\frac{1}{2}\left(\frac{M_{1}}{x_{1}}+\frac{M_{2}}{x_{2}}\right)+
\frac{1}{2}\left(\frac{x_{1}}{m_{1}}+\frac{x_{2}}{m_{2}}\right)\right\}.
\end{eqnarray*}
Let $g_{1}(x_{1})=\frac{M_{1}}{x_{1}}-\frac{x_{1}}{m_{1}}$ and 
$g_{2}(x_{2})=-(\frac{M_{2}}{x_{2}}-\frac{x_{2}}{m_{2}})$. Then 
$(p_{1},p_{2}) \in {\cal S}_{f_{2}}^{2}$ iff $g_{1}(p_{1})=g_{2}(p_{2})$. Notice that 
$g_{1}$ is monotonically decreasing on $[m_{1},M_{1}]$ 
and $g_{2}$ is monotonically increasing on $[m_{2},M_{2}]$. Then for any 
$x_{1} \in [m_{1},M_{1}]$, we have that 
\[
- \left(\frac{M_{1}}{m_{1}}-1\right) =
g_{1}(M_{1}) \leq g_{1}(x_{1}) \leq g_{1}(m_{1})=\frac{M_{1}}{m_{1}}-1,
\]
and for any $x_{2} \in [m_{2},M_{2}]$, we also have that 
\[
-\left(\frac{M_{2}}{m_{2}}-1\right) =
g_{2}(m_{2}) \leq g_{2}(x_{2}) \leq g_{2}(M_{2})=\frac{M_{2}}{m_{2}}-1.
\]
For any $(x_{1},x_{2}) \in {\cal S}_{f_{2}}^{2}$, we 
claim that $-(\frac{M_{2}}{m_{2}}-1) \leq g_{1}(x_{1}) \leq \frac{M_{2}}{m_{2}}-1$\footnote{~Recall that  $-(\frac{M_{2}}{m_{2}}-1) \leq g_{2}(x_{2}) \leq 
\frac{M_{2}}{m_{2}}-1$. If 
$-(\frac{M_{2}}{m_{2}}-1) > g_{1}(x_{1})$ or $\frac{M_{2}}{m_{2}}-1< g_{1}(x_{1})$, 
then $(x_{1},x_{2}) \not \in {\cal S}_{f_{2}}$.}. 
Let $L_{1} \in [m_{1},M_{1}]$~such~that 
$g_{1}(L_{1})=g_{2}(M_{2})=\frac{M_{2}}{m_{2}}-1$ and 
$R_{1} \in [m_{1},M_{1}]$ such that 
$g_{1}(R_{1})=g_{2}(m_{2})=-(\frac{M_{2}}{m_{2}}-1)$,~i.e., 
\begin{eqnarray*}
L_{1} & = & - \frac{m_{1}}{2}\left(\frac{M_{2}}{m_{2}}-1\right)+
\sqrt{\left\{\frac{m_{1}}{2} 
\left(\frac{M_{2}}{m_{2}}-1\right)\right\}^{2}+m_{1}M_{1}};\\
R_{1} & = & \frac{m_{1}}{2}\left(\frac{M_{2}}{m_{2}}-1\right)+
\sqrt{\left\{\frac{m_{1}}{2} 
\left(\frac{M_{2}}{m_{2}}-1\right)\right\}^{2}+m_{1}M_{1}}.
\end{eqnarray*}
It is immediate that $(L_{1},M_{2}) \in {\cal S}_{f_{2}}^{2}$ 
and $(R_{1},m_{2}) \in {\cal S}_{f_{2}}^{2}$. 

Let $h_{1}(x_{1})=\frac{1}{2}(\frac{M_{1}}{x_{1}}+\frac{x_{1}}{m_{1}})$ and 
$h_{2}(x_{2})=\frac{1}{2}(\frac{M_{2}}{x_{2}}+\frac{x_{2}}{m_{2}})$. 
Since $h_{1}$ is convex on $[L_{1},R_{1}]\subseteq [m_{1},M_{1}]$ 
and $h_{2}$ is convex on $[m_{2},M_{2}]$,  we have that 
$\max_{x_{1}\in [L_{1},R_{1}]} h_{1}(x_{1}) = 
\max\{h_{1}(L_{1}),h_{1}(R_{1})\}$, where 
\[
h_{1}(L_{1}) = h_{1}(R_{1})=
\sqrt{\left\{\frac{1}{2}\left(\frac{M_{2}}{m_{2}}-1\right)\right\}^{2}
+ \frac{M_{1}}{m_{1}}}, 
\]
and $\max_{x_{2}\in [m_{2},M_{2}]} h_{2}(x_{2}) = 
\max\{h_{2}(m_{2}),h_{2}(M_{2})\}$, where $h_{2}(m_{2})=h_{2}(M_{2})=
\frac{1}{2}(\frac{M_{2}}{m_{2}}+1)$.~Thus~it follows that $z_{f_{2}}^{2} = 
\frac{1}{2}\{h_{1}(L_{1})+h_{2}(M_{2})\}
=\frac{1}{2}\{h_{1}(R_{1})+h_{2}(m_{2})\}$. \BQED\medskip

With respect to the function $f_{2}$ for $k=2$, 
Tiedemann, et al. \cite{TIS}  presented the algorithm {\sc rpp-mult} 
and showed that 
${\cal CR}^{f_{2}}(\msc{rpp-mult}) \leq 
\sqrt[4]{(M_{1}M_{2})/(m_{1}m_{2})}$ \cite[Theorem 3]{TIS} 
(this~is~shown~by Definition \ref{df-ratio}, 
but also can be shown by Definition \ref{df-mod-ratio}). Note 
that $\sqrt[4]{(M_{1}M_{2})/(m_{1}m_{2})} <z_{f_{2}}^{2}$.~So from 
Theorems 
\ref{thm-unified-lower} and \ref{thm-arithmetic}, we have that 
${\cal CR}^{f_{2}}(\msc{alg}_{2}) \geq z_{f_{2}}^{2}$ for 
any algorithm $\msc{alg}_{2}$,~which~disproves the result 
\cite[Theorem 3]{TIS}. 
This is because in the proof of the result 
\cite[Theorem 3]{TIS}, the maximum in \cite[Equation (9)]{TIS} 
cannot be achieved 
at $\sqrt{M_{1} z^{*}/M_{2}}$, where 
$z^{*}=\sqrt{m_{1}M_{2}m_{2}M_{1}}$. 
%
%=========================================================
\subsection{Geometric Mean Component Competitive Ratio} 
\label{subsec-geometric}
%=========================================================
%
For $c_{1},\ldots,c_{k} \in {\bf R}$, 
let $f_{3}(c_{1},\ldots,c_{k})=\sqrt[k]{\prod_{i=1}^{k}c_{i}}$. 
For the function 
$f_{3}: {\bf R}^{k} \to {\bf R}$, let 
\begin{eqnarray*}
{\cal S}_{f_{3}}^{k} & = & \left\{(x_{1},\ldots,x_{k})\in I_{1} \times 
\cdots \times I_{k}: \sqrt[k]{\prod_{i=1}^{k} \frac{M_{i}}{x_{i}}} = 
\sqrt[k]{\prod_{i=1}^{k} \frac{x_{i}}{m_{i}}} \right\};\\
z_{f_{3}}^{k} & = & \sup_{(x_{1},\ldots,x_{k}) \in  {\cal S}_{f_{3}}^{k}} 
\sqrt[k]{\prod_{i=1}^{k} \frac{M_{i}}{x_{i}}}.
\end{eqnarray*}
With respect to the function $f_{3}$ for $k=2$, 
it is easy to see that 
the algorithm $\msc{bpp}_{2}$ is identical to the algorithm 
$\msc{rpp-mult}$ \cite{TIS}. In fact, 
Tiedemann, et al. \cite{TIS} showed that 
${\cal CR}^{f_{3}}(\msc{rpp-mult}) = 
\sqrt[4]{(M_{1}M_{2})/(m_{1}m_{2})}$ 
with respect to 
the function $f_{3}$ for $k=2$, and 
this can be generalized~to~the result that 
${\cal CR}_{s}^{f_{3}}(\msc{bpp}_{k})=z_{f_{3}}^{k}$ for any integer $k \geq 2$ 
(see Corollary \ref{cor-unified} with respect to $f_{3}$). 
\begin{theorem} \label{thm-geometric}
$z_{f_{3}}^{k}=\sqrt[2k]{\prod_{i=1}^{k} M_{i}/m_{i}}$ 
for any integer $k \geq 2$. 
\end{theorem}
{\bf Proof:} From the definition of ${\cal S}_{f_{3}}^{k}$, it follows that 
$\sqrt[k]{\prod_{i=1}^{k} M_{i}/x_{i}}=
\sqrt[k]{\prod_{i=1}^{k} x_{i}/m_{i}}$ 
for any integer $k \geq 2$ and any 
$(x_{1},\ldots,x_{k}) \in {\cal S}_{f_{3}}^{k}$. Then 
$\prod_{i=1}^{k} x_{i} = \sqrt{\prod_{i=1}^{k} m_{i}M_{i}}$, and this implies that 
\[
\prod_{i=1}^{k} \frac{M_{i}}{x_{i}} = \frac{\prod_{i=1}^{k} M_{i}}{\prod_{i=1}^{k} x_{i}}
= \frac{\prod_{i=1}^{k} M_{i}}{\sqrt{\prod_{i=1}^{k} m_{i}M_{i}}} 
= \sqrt{\prod_{i=1}^{k} \frac{M_{i}}{m_{i}}}. 
\]
Thus we can conclude that 
$z_{f_{3}}^{k} = \sup_{(x_{1},\ldots,x_{k}) \in  {\cal S}_{f_{3}}^{k}} 
\sqrt[k]{\prod_{i=1}^{k} M_{i}/x_{i}} = 
\sqrt[2k]{\prod_{i=1}^{k} M_{i}/m_{i}}$. \BQED
%
%=========================================================
\subsection{Best Component Competitive Ratio} 
\label{subsec-best}
%=========================================================
%
In this subsection, we deal with a new and natural 
continuous monotone function $f_{4}: {\bf R}^{k}\to {\bf R}$. 
For $c_{1},\ldots,c_{k} \in {\bf R}$, 
let $f_{4}(c_{1},\ldots,c_{k})=\min(c_{1},\ldots,c_{k})$. 
For the function $f_{4}: {\bf R}^{k} \to {\bf R}$, let 
\begin{eqnarray*}
{\cal S}_{f_{4}}^{k} & = & \left\{(x_{1},\ldots,x_{k})\in 
I_{1} \times \cdots \times I_{k}: 
\min\left(\frac{M_{1}}{x_{1}},\ldots,\frac{M_{k}}{x_{k}}\right)=
\min\left(\frac{x_{1}}{m_{1}},\ldots,\frac{x_{k}}{x_{k}}\right)\right\};\\
z_{f_{4}}^{k} & = & \sup_{(x_{1},\ldots,x_{k}) \in  {\cal S}_{f_{4}}^{k}} 
\left[ \min\left(\frac{M_{1}}{x_{1}},\ldots,\frac{M_{k}}{x_{k}}\right)\right]. 
\end{eqnarray*}
\begin{theorem} \label{thm-best}
$z_{f_{4}}^{k} = \sqrt{M_{k}/m_{k}}$ 
for any integer $k \geq 1$. 
\end{theorem}
{\bf Proof:} We first show that $z_{f_{4}}^{k} \leq \sqrt{M_{k}/m_{k}}$. 
Assume by contradiction that $z_{f_{4}}^{k} > \sqrt{M_{k}/m_{k}}$ and 
let $\vec{y} = (y_{1},\ldots,y_{k}) \in {\cal S}_{f_{4}}^{k}$ 
such that 
\[
z_{f_{4}}^{k} =f_{4}\left(\frac{M_{1}}{y_{1}},\ldots,
\frac{M_{k}}{y_{k}}\right) = 
f_{4}\left(\frac{y_{1}}{m_{1}},\ldots,
\frac{y_{k}}{m_{k}} \right ) > 
\sqrt{\frac{M_{k}}{m_{k}}}.
\]
Since $f_{4}(c_{1},\ldots,c_{k})=\min(c_{1},\ldots,c_{k})$, we have that 
for each $i \in [1,k]$, 
\begin{eqnarray*}
\frac{M_{i}}{y_{i}} & \geq & z_{f_{4}}^{k} > \sqrt{\frac{M_{k}}{m_{k}}};\\
\frac{y_{i}}{m_{i}} & \geq & z_{f_{4}}^{k} > \sqrt{\frac{M_{k}}{m_{k}}}.
\end{eqnarray*}
In particular, we have that $M_{k}/y_{k} > \sqrt{M_{k}/m_{k}}$ and 
$y_{k}/m_{k} > \sqrt{M_{k}/m_{k}}$. This implies that 
\[
\frac{M_{k}}{m_{k}} = \frac{M_{k}}{y_{k}} \cdot \frac{y_{k}}{m_{k}} 
> \sqrt{\frac{M_{k}}{m_{k}}}\cdot \sqrt{\frac{M_{k}}{m_{k}}} 
= \frac{M_{k}}{m_{k}}, 
\]
and this is a contradiction. So 
it follows that $z_{f_{4}}^{k} \leq \sqrt{M_{k}/m_{k}}$. 
Next we show that 
there~exists $\vec{x}^{*}=(x_{1}^{*},\ldots,x_{k}^{*}) \in {\cal S}_{f_{4}}^{k}$ 
such that 
\[
z_{f_{4}}^{k} =f_{4}\left(\frac{M_{1}}{x_{1}^{*}},\ldots,
\frac{M_{k}}{x_{k}^{*}}\right) = 
f_{4}\left(\frac{x_{1}^{*}}{m_{1}},\ldots,
\frac{x_{k}^{*}}{m_{k}} \right ) = \sqrt{\frac{M_{k}}{m_{k}}}.
\]
For each $i \in [1,k]$, let $x_{i}^{*}=\sqrt{m_{i}M_{i}}$. Then it is immediate that 
\begin{eqnarray*}
f_{4}\left(\frac{M_{1}}{x_{1}^{*}},\ldots,\frac{M_{k}}{x_{k}^{*}}\right)
& = & \min\left\{\frac{M_{1}}{\sqrt{m_{1}M_{1}}}, \ldots,
\frac{M_{k}}{\sqrt{m_{k}M_{k}}} \right\}\\
& = & \min\left\{\sqrt{\frac{M_{1}}{m_{1}}}, \ldots,
\sqrt{\frac{M_{k}}{m_{k}}} \right\}=\sqrt{\frac{M_{k}}{m_{k}}};\\
f_{4}\left(\frac{x_{1}^{*}}{m_{1}},\ldots,\frac{x_{k}^{*}}{m_{k}}\right)
& = & \min\left\{\frac{\sqrt{m_{1}M_{1}}}{m_{1}}, \ldots,
\frac{\sqrt{m_{k}M_{k}}}{m_{k}} \right\}\\
& = & \min\left\{\sqrt{\frac{M_{1}}{m_{1}}}, \ldots,
\sqrt{\frac{M_{k}}{m_{k}}} \right\}=\sqrt{\frac{M_{k}}{m_{k}}}.
\end{eqnarray*}
Thus we have that $z_{f_{4}}^{k}=\sqrt{M_{k}/m_{k}}$ for each integer 
$k \geq 1$. \BQED
%
%=================================================
\section{Concluding Remarks} \label{sec-remark}
%=================================================
%
In this paper, we have proposed a simple online algorithm Balanced Price Policy 
($\msc{bpp}_{k}$) for the multi-objective ($k$-objective) 
time series search problem and have shown that $\msc{bpp}_{k}$ 
is {\it best possible\/} with respect to any monotone (not necessarily 
continuous) function $f: {\bf R}^{k} \to {\bf R}$ even if all of 
$\msc{itv}_{1}=[m_{1},M_{1}],\ldots,\msc{itv}_{k}=[m_{k},M_{k}]$ are not 
necessarily real intervals 
(Theorem \ref{thm-general}). 
In the case that 
all of $\msc{itv}_{1}=[m_{1},M_{1}],\ldots,\msc{itv}_{k}=[m_{k},M_{k}]$ 
are real intervals, we have formulated the best possible value of the 
competitive ratio 
exactly for any monotone continuous function 
$f: {\bf R}^{k}\to {\bf R}$ (Theorems \ref{thm-unified-upper} 
and \ref{thm-unified-lower}). 
We also have derived the best possible~values~of~the~competitive ratio 
for the multi-objective time series search problem with respect to 
several known measures of the competitive analysis, i.e., 
the best possible value of 
the competitive ratio for the multi-objective time series search problem with 
respect to the worst component competitive analysis 
(Theorem \ref{thm-worst}), 
the best possible value of the competitive ratio 
for the bi-objective time 
series search problem with respect to the 
arithmetic mean component competitive analysis 
(Theorem \ref{thm-arithmetic}), and 
the best possible value of 
the competitive ratio for the multi-objective time series search 
problem with respect to the geometric mean component competitive analysis 
(Theorem \ref{thm-geometric}). For a new measure of the competitive analysis, 
we derive the best possible value of the competitive ratio 
for the multi-objective time series search problem with 
respect to the best component competitive analysis (Theorem \ref{thm-best}). 

For each $i \in [1,k]$, let $I_{i}=[m_{i},M_{i}]$ with $0<m_{i}\leq M_{i}$.  
%Let $f_{2}(x_{1},\ldots,x_{k}) = (x_{1}+\cdots+x_{k})/k$. 
As we have shown in Theorem~\ref{thm-arithmetic}, 
the best possible value of the competitive ratio for the bi-objective time series 
search problem with respect to the 
arithmetic mean component competitive analysis is 
\[
z_{f_{2}}^{2}=\frac{1}{2}\left[\sqrt{
\left\{\frac{1}{2}
\left(\frac{M_{2}}{m_{2}}-1\right)\right\}^{2} 
+ \frac{M_{1}}{m_{1}}}+ \frac{1}{2}\left(\frac{M_{2}}{m_{2}}+1\right)\right].
\]
In Corollary \ref{cor-unified}, we have given 
the best possible value $z_{f_{2}}^{k}$ of the competitive 
ratio~for~the~multi-objective ($k$-objective) 
time series search problem with respect to the 
arithmetic mean component competitive analysis, where
\begin{eqnarray*}
z_{f_{2}}^{k} & = & \sup_{(x_{1},\ldots,x_{k}) \in {\cal S}_{f_{2}}^{k}}
f_{2}\left(\frac{M_{1}}{x_{1}},\ldots,\frac{M_{k}}{x_{k}}\right),\\
{\cal S}_{f_{2}}^{k} & = & \left\{(x_{1},\ldots,x_{k})\in I_{1} \times 
\cdots \times I_{k}: f_{2}\left(\frac{M_{1}}{x_{1}}, \ldots,
\frac{M_{k}}{x_{k}}\right)=
f_{2}\left(\frac{x_{1}}{m_{1}}, \ldots,\frac{x_{k}}{m_{k}}\right)\right\}. 
\end{eqnarray*}

So we have the following interesting open 
problem for the multi-objective time series search problem with respect to 
the arithmetic mean component competitive analysis. 
\begin{namelist}{~~~(2)}
\item[(1)] For any integer $k \geq 3$, find an explicit representation of 
$z_{f_{2}}^{k}$ or 
natural conditions for 
$m_{1},\ldots,m_{k},M_{1},\ldots,M_{k}$ to  explicitly represent $z_{f_{2}}^{k}$. 
\end{namelist}

In fact, we may have many practical multi-objective online 
problems other than the multi-objective time series search problem. Then 
we also have the following problem for future work. 
\begin{namelist}{~~~(2)}
\item[(2)] For a practical multi-objective ($k$-objective) 
online problem ${\cal P}_{k}$, design an efficient online algorithm 
$\msc{alg}_{k}$ with respect to a natural monotone 
function $f:{\bf R}^{k}\to {\bf R}$, and analyze the competitive ratio 
of the algorithm $\msc{alg}_{k}$ with respect to the monotone function 
$f$. 
\end{namelist}
%
%===========================

%====================
%
\end{document}